\documentclass{aa}  

\usepackage{graphicx}
\usepackage{txfonts}
\usepackage{lipsum}
\usepackage{subcaption}
\usepackage{lscape}    
                               
\usepackage{placeins}           

\usepackage{hyperref}

\begin{document}

   \title{Dynamical evolution of the stellar mass function in multiple-population globular clusters}

   \author{J. Bruce\inst{1}\fnmsep\thanks{E-mail: jobruce@iu.edu}
        \and E. Vesperini\inst{1}
        \and A. Askar\inst{2}
        \and M. Giersz\inst{2}
        \and E. Dalessandro\inst{3}
        \and M. Scalco\inst{1}
        }

   \institute{Department of Astronomy, Indiana University, Swain West, 727 E. 3rd Street, Bloomington, IN 47405, USA
   \and
   Nicolaus Copernicus Astronomical Center, Polish Academy of Sciences, ul. Bartycka 18, Warsaw 00--716, Poland
   \and 
   INAF, Osservatorio di Astrofisica e Scienza dello Spazio di Bologna, Via Gobetti 93/3, Bologna 40129, Italy}

   \date{Submitted --}
 
  \abstract
    {\noindent The dynamical evolution of globular clusters is significantly influenced by the multiscale structure introduced by the presence of multiple stellar populations. In this study, we investigate the long-term dynamical evolution of the stellar mass function (MF) within globular clusters hosting two stellar populations using a suite of Monte Carlo simulations run with the MOCCA code. Dynamical processes including two-body relaxation, mass segregation, and tidal stripping result in distinct radial profiles describing the variation with distance from the cluster's center of the stellar MF slope within the first population (P1), the second population (P2), and the cluster as a whole. We show that these radial profiles are well described by a broken power law with a knee located slightly interior to the half-mass radius, a feature partially driven by the multiscale structure of the multiple populations. In both 3D and 2D projected coordinates, we explore several characteristic clustercentric distances within the profile, including those where the local MF slope matches the global value of the present-day MF slope and the initial mass function (IMF) slope, and we examine how these positions evolve with the cluster's dynamical age. Our results highlight how the initially more centrally concentrated P2 drives a steeper radial gradient and a steeper MF slope in the outer regions, which subsequently affects the global cluster values. Motivated by recent observational studies suggesting that a bottom-light IMF may better describe massive globular clusters, as well as measurements of a flatter P2 MF slope in the outer regions of 47 Tucanae, we additionally explore a three-parameter bottom-light IMF and a simulation in which the P1 IMF differs from the P2 IMF, with the P2 IMF flatter than that of the P1 in the lower mass range.}

    \keywords{globular clusters: general, mass function}

   \maketitle
   \nolinenumbers

\section{Introduction}

Galactic globular clusters (GCs) are among the oldest stellar systems known, with ages often exceeding 12 Gyr. This makes them excellent laboratories for studying stellar populations that formed in the early universe and have since been subjected to long-term dynamical evolution within an active stellar environment. However, this same dynamical complexity makes their evolution difficult to disentangle, as numerous interacting processes shape their present-day properties. 

One active area of research concerns the evolution of multiple stellar populations (MPs), which are predicted by many theoretical models to form with a centrally concentrated second population (P2) embedded within a more extended first population (P1). {Differences between the P2 and the P1 spatial distributions consistent with the theoretical predictions have been observed in several clusters \citep[see e.g.][]{Bellini+2009, Lardo+2011, Beccari+2013, Cordero+2014, simioni+2016, Boberg+2016, Dalessandro+2019, Onorato+2023, Cadelano+2024}. The distinct dynamical environments and multiscale structure introduced by the presence of MPs have significant implications for the evolution of many cluster properties, including binary stars \citep[see e.g.][]{Vesperini+2011, Hong+2015, Lucatello+2015, Hong+2016, Dalessandro+2018b, Hong+2019, Hypki+2022, Bortolan+2025, Bruce+2026, Bruce+2026b}, and the stellar mass function (MF) \citep[][]{Vesperini+2018}. The formation of P2 has been variously attributed to the reaccretion of pristine gas with pollution from sources such as AGB stars, supermassive stars, or massive binaries \citep[see e.g.][]{Ventura+2001, deMink+2009, Bastian+2013, Gieles+2018, Calura+2019, Nguyen+Sills2024}.

The distribution of stellar masses within a cluster is commonly described by a stellar MF, which evolves from an initial mass function (IMF) imprinted during the star formation phase. The dynamical evolution of the stellar MF and its radial variation within a cluster depend on many processes, including the evaporation of stars through two-body relaxation, ejection through encounters in the core, and tidal stripping from the outermost regions. Previous theoretical studies have predicted the formation of radial gradients in the MF slope, where the inner regions become flatter through mass segregation, while the outer regions become steeper as lower-mass stars migrate outward \citep[see e.g.][]{Vesperini+Heggie1997, Baumgardt+makino2003, Webb+vesperini2016, Webb+2017}; similar trends have been identified in several observations studies \citep[see e.g.][]{Beccari+2011, Dalessandro+2015}. The presence of MPs is expected to affect the formation of this radial gradient, and the subsequent preferential loss of lower-mass stars from the outer regions, in ways that depend on the unique dynamical environment of each population.

The assumption of a universal IMF has been recently challenged by observational studies finding results inconsistent with a \citet{Kroupa2001} IMF \citep[see e.g.][Dalessandro et al. in prep]{vandokkum2008, Giersz+2011, Webb+2017, Cadelano+2020, Baumgardt+2023, Dickson+2023}. Additionally, unique observational signatures within MP studies seem to be indicative of more complex IMF prescriptions, possibly hinting at distinct IMFs associated with the chemically and dynamically distinct MPs \citep{Scalco+2026}.

In this paper we investigate the long-term dynamical evolution of the stellar MF in GCs hosting multiple stellar populations using a suite of MOCCA (where MOCCA stands for MOnte Carlo Cluster simulAtor) simulations \citep[][]{Giersz1998, Hypki+2013, Hypki+2022, Hypki+2025}. We detail how the multiscale structure and the differing dynamical environments within each population uniquely drive the evolution of the MF, resulting in distinct global values and radial profiles of the variation of the local MF slope with the clustercentric distance. We further show that the radial profiles can be described by a broken power law that evolves over time, and we characterize the evolution of several radial features, including the knee of this broken power law, the location where the local MF slope is equal to the global value of the present-day MF, and the location where the local MF slope is equal to the IMF value.

The paper is organized as follows. In Section~\ref{sec:2} we describe the setup of the simulations used in this work. Section~\ref{sec:3} presents our results: Section~\ref{sec:3.1} examines the evolution of the radial profiles, Section~\ref{sec:3.2} explores the evolution of the global MF properties, Section~\ref{sec:3.3} compares the local and global slopes of the MF, and Section~\ref{sec:3.4} discusses the differences introduced by varying the IMFs within the populations. Finally, Section~\ref{sec:4} summarizes and discusses the main findings of this paper. In the appendices we present additional results for different mass ranges (Appendix~\ref{sec:A1}) and observational projections (Appendix~\ref{sec:A2}).

\section{Methods}
\label{sec:2}

This study investigates the dynamical evolution of GCs using the MOCCA code \citep[][]{Giersz1998, Hypki+2013}. This Monte Carlo-based code incorporates stellar evolution through the SSE and BSE algorithms (\citet{Hurley+2000, Hurley+2002}, with additional updates described in \citet{kamlah+2022}), and efficiently models two-body relaxation and the long-term dynamical evolution of GCs.

The initial conditions for the simulations presented in this work are similar to those used in previous studies investigating the long-term dynamical evolution of star clusters hosting MPs, with both populations initially present \citep[see e.g.][]{Hong+2015, Hong+2016, Vesperini+2021, Hypki+2022, Bruce+2026, Bruce+2026b}. Each simulation begins with a total of one million stars, divided between an extended P1 and a more centrally concentrated P2, with either 10\% or 25\% of stars initially belonging to P2. The centrally concentrated P2 setup is motivated by the results of hydrodynamical simulations of MP formation \citep[see e.g.][]{Dercole+2008, Bekki2010, Bekki2011, Calura+2019}. The ratio of the P2 to P1 half-mass radii is initialized such that P2 is either 10 or 20 times more centrally concentrated than the tidally filling P1.

Each simulation begins with an initial binary fraction of 10\% within each population. For the calculation of the slope of the mass function we exclude binary stars and use single stars only (including all binaries as unresolved sources with the combined mass of their two components changes the measured slopes by less than 0.2 within both populations across all clustercentric distances considered in our analysis, and including binaries as resolved sources and adding each individual binary component as a single star has an even smaller effect). The initial binary properties (binary pairing, mass ratio and semi-major axis distribution) are the same as those adopted in \citet{Bruce+2026} and we refer to that paper for further details. In the models explored in this work, we adopt the same natal kicks for neutron stars and black holes, with velocities following a Maxwellian distribution with a 1-D velocity dispersion of $\sigma=265$ km/s \citep[][]{hobbs+2005}. As a result, the majority of black holes are ejected shortly after formation due to natal kicks (with $\sim 0-3$ black holes remaining after 1 Gyr). The presence of a significant fraction of black holes could reduce the degree of mass segregation of the visible stars as well as have other effects on the cluster evolution. We plan to explore this in a future paper.

The metallicity is initially set to $Z = 0.001$, and the tidal radius is set to 59.1 pc (consistent, for example, with a cluster moving on a circular orbit at a galactocentric distance of 3.3 kpc in a point-mass galactic potential). The cluster is initially in virial equilibrium, with the two populations modeled by King distributions \citep{King1966} with central concentration parameters $W_0=5$ for P1 and $W_0=7$ for P2. We also include a simulation run with a smaller initial tidal radius of 38.0 pc (analogous, for example, to a cluster orbiting at a galactocentric distance of 1.7 kpc); the shorter relaxation time and stronger tidal field of this system results in enhanced stellar escape, allowing the cluster to reach an older dynamical age by the end of the simulation. Additionally, we run two single population (SP) simulations that are either tidally underfilling or tidally filling. A summary of the initial properties of all simulations is provided in Table~\ref{tab:tab 1}.

\begin{table}[ht]
\centering
\caption{Initial simulation parameters.}
\begin{tabular}{lcccc}
\hline
\hline
Simulation & $F_{P2}$ & $r_{h, P2}/r_{h, P1}$ & Initial $r_t$ (pc) & $r_{hm}/r_t$\\
\hline
mr025c01 & 25\% & 0.1 & 59.1 & 0.14\\
mr025c005 & 25\% & 0.05 & 59.1 & 0.14\\
mr01c01 & 10\% & 0.1 & 59.1 & 0.17\\
mr025c005sf & 25\% & 0.05 & 38.0 & 0.14\\
3pl & 25\% & 0.05 & 68.5 & 0.14\\
3pl-anom & 25\% & 0.05 & 68.5 & 0.13\\
SPtf & - & - & 59.1 & 0.12\\
SPuf & - & - & 362.2 & 0.003\\
\hline
\end{tabular}
\tablefoot{Initially, the metallicity (Z) of all simulations is 0.001, and we define $F_{P2}$ as the fraction of total objects (single or binary stars) in the P2 population. The initial mass of all simulations is $6.378\times10^5\,M_\odot$, except for the 3pl ($M_{ini} = 1.0102\times10^6\,M_\odot$) and 3pl-anom ($M_{ini} = 9.91\times10^5\,M_\odot$) models.}
\label{tab:tab 1}
\end{table}

We explored both two- and three-parameter IMFs, with initial stellar masses spanning between $0.1$ -- $100\,M_\odot$. In all models, the P1 and P2 IMFs share the same maximum stellar mass of $100M_\odot$; we discuss the effect of adopting a lower P2 maximum mass in Section~\ref{sec:3.4}. As a standard two-parameter IMF, we adopt the \citet{Kroupa2001} IMF, hereafter referred to as K01, which is characterized by a slope of $-1.3$ in the mass range $0.1$--$0.5\,M_\odot$, and $-2.3$ in the mass range $0.5$--$100\,M_\odot$. We also investigate a three-parameter MF with slopes suggested in \citet{Baumgardt+2023}, hereafter referred to as 3pl, as well as a modified version of the 3pl model (hereafter referred to as 3pl-anom) in which the P1 and P2 IMFs have different slopes in the low-mass range, motivated by a recent observational study of \citet{Scalco+2026} based on JWST data and suggesting possible differences between the P1 and P2 IMF in 47 Tuc. A summary of the IMF slopes within the different mass ranges for the 3pl and 3pl-anom simulations is provided in Table~\ref{tab:tab 2}. Throughout the paper, we report the values of the MF slope for the K01 simulations for main sequence stars in the mass range $0.1-0.5\,M_\odot$, and for the 3pl and 3pl-anom simulations for main sequence stars within $0.1-0.4\,M_\odot$, matching the lowest IMF bin boundary of the models (see Table~\ref{tab:tab 2}). We verified that adopting a narrower range ($0.1-0.4\,M_\odot$) for the K01 simulations produced consistent results, so this is not expected to affect the comparisons between the K01 and 3pl simulations. The motivation for the 3pl model comes from observational studies that suggest a bottom-light IMF better describes the values measured in massive GCs \citep[][]{Giersz+2011, Webb+2017, Cadelano+2020, Baumgardt+2023}, while the 3pl-anom model is motivated by the observations of a flatter MF slope in P2 compared to P1 for 47 Tucanae \citep{Scalco+2026}.

\begin{table}[ht]
\centering
\caption{Initial 3pl simulation mass functions.}
\begin{tabular}{lccc}
\hline
\hline
Mass range ($M_\odot$) & Population & 3pl & 3pl-anom \\
\hline
$0.1\leq M <0.4$ & P1 & -0.3  & -0.5  \\
$0.4\leq M <1.0$ & P1 & -1.65 & -1.65 \\
$1.0\leq M <100$ & P1 & -2.3  & -2.3  \\
$0.1\leq M <0.4$ & P2 & -0.3  & -0.05 \\
$0.4\leq M <1.0$ & P2 & -1.65 & -1.65 \\
$1.0\leq M <100$ & P2 & -2.3  & -2.3  \\
\hline
\end{tabular}
\label{tab:tab 2}
\end{table}

We characterize the stellar MF by fitting a single power law of the form $\zeta(m) \propto m^\alpha$, where $\zeta(m)=dN/dm$ is the number of stars per unit mass and $\alpha$ is the MF slope. We determined $\alpha$ by maximizing the likelihood function outlined in \citet{Trenti+2010}, which is evaluated directly on the individual stellar mass of single main-sequence stars in the sample, without any prior binning. We maximized the likelihood using a Markov Chain Monte Carlo (MCMC) approach, where we report the median of the posterior with error bars or shaded regions highlighting the 16th and 84th percentiles. We emphasize that although our choice of initial conditions is not designed to model specific Galactic clusters or to span the full range of observed properties, the final values of some key properties of our models are broadly consistent with those observed in many Milky Way GCs \citep[see][for the final mass, P2 fraction, and $R_{hl}$ of our models]{Bruce+2026}.

\section{Results}
\label{sec:3}

\begin{figure}[h!]
    \centering
    \includegraphics[width=0.95\linewidth]{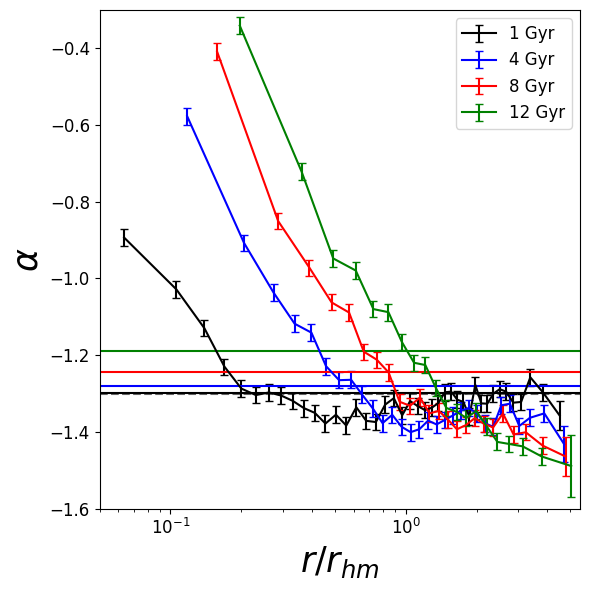}
    \caption{The slope of the stellar MF ($\alpha$) as a function of distance from the cluster center, normalized to the half-mass radius, at four different time snapshots for the mr025c005 simulation, within the mass range $0.1$--$0.5\,M_\odot$. Each point represents the median stellar position within bins of equal stellar number. The horizontal lines display the global value of $\alpha$ at the different times considered in this figure. The horizontal gray dashed line displays the initial value of $\alpha$.}
    \label{fig:fig 1}
\end{figure}

\begin{figure}[h!]
    \centering
    \includegraphics[width=0.95\linewidth]{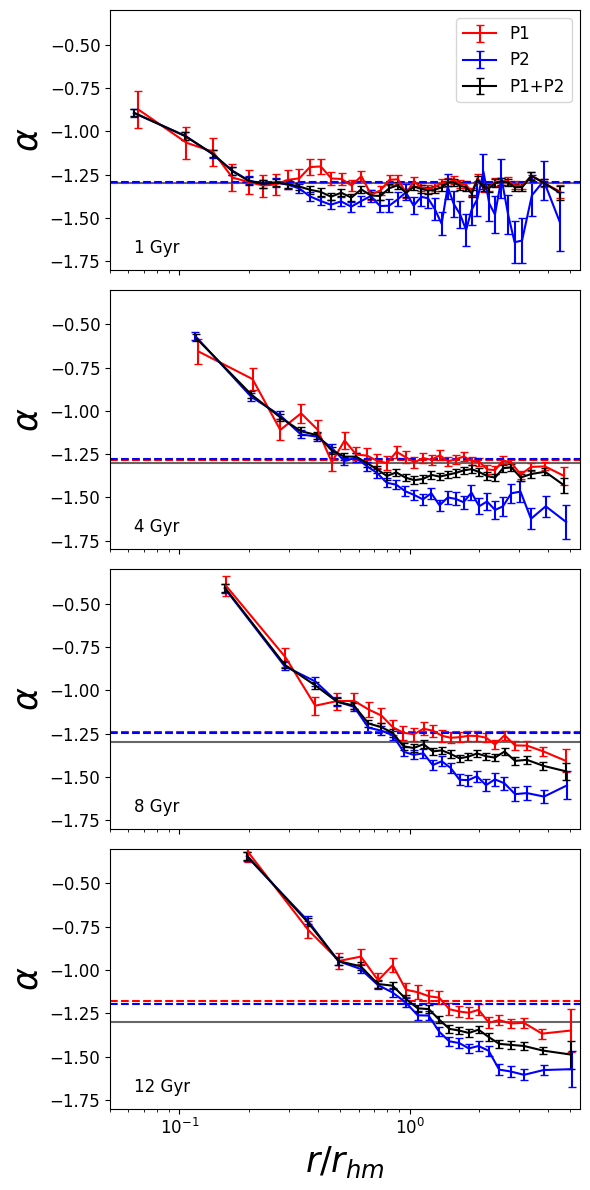}
    \caption{The slope of the stellar MF ($\alpha$) as a function of distance from the cluster center, normalized to the half-mass radius, at 1, 4, 8, and 12 Gyr snapshots within P1 (red), P2 (blue), and P1+P2 (black) for the mr025c005 simulation, within the mass range $0.1$--$0.5\,M_\odot$. The horizontal lines display the current global value of $\alpha$ within each population. The horizontal gray line displays the initial value of $\alpha$.}
    \label{fig:fig 2}
\end{figure}

\begin{figure}[h]
    \centering
    \includegraphics[width=0.95\linewidth]{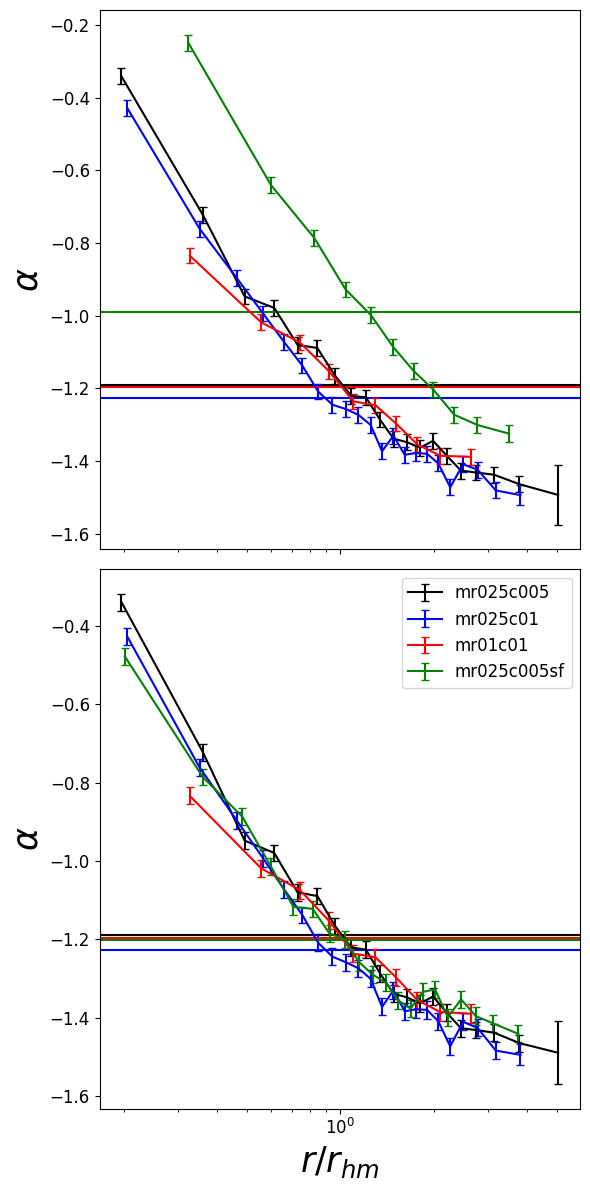}
    \caption{The slope of the stellar MF ($\alpha$) as a function of distance from the cluster center, normalized to the half-mass radius, for four simulations at 12 Gyr (top panel), and at a dynamical age $t/t_{rh}(t)\sim5$ (bottom panel), within the mass range $0.1$--$0.5\,M_\odot$. The horizontal lines display the current global value of $\alpha$.}
    \label{fig:fig 3}
\end{figure}

\begin{figure}[h]
    \centering
    \includegraphics[width=0.95\linewidth]{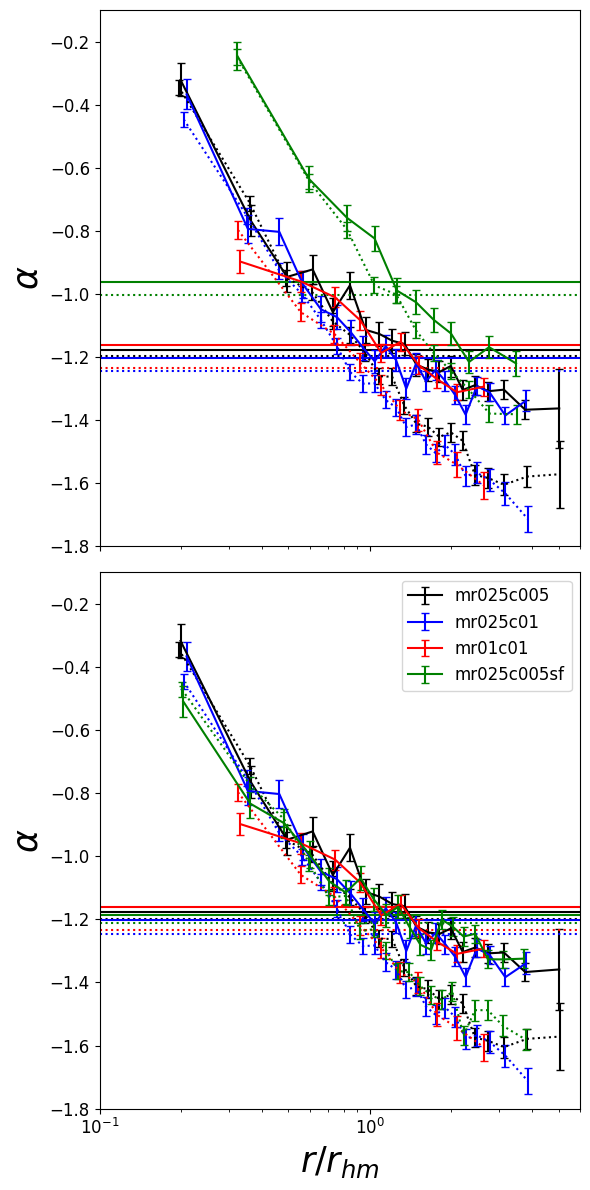}
    \caption{The slope of the stellar MF ($\alpha$) as a function of distance from the cluster center, normalized to the half-mass radius, for each simulation at 12 Gyr (top panel), and at a dynamical age $t/t_{rh}(t)\sim5$ (bottom panel), within P1 (solid lines) and P2 (dotted lines), within the mass range $0.1$--$0.5\,M_\odot$. The horizontal lines display the current global value of $\alpha$ within each population.}
    \label{fig:fig 4}
\end{figure}

\begin{figure}[h]
    \centering
    \includegraphics[width=0.95\linewidth]{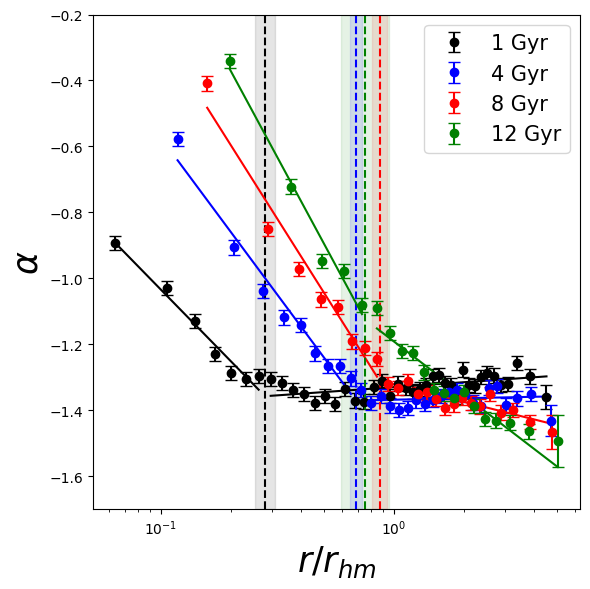}
    \caption{The slope of the stellar MF ($\alpha$) as a function of distance from the cluster center, normalized to the half-mass radius, at four different time snapshots for the mr025c005 simulation, within the mass range $0.1$--$0.5\,M_\odot$. The vertical dashed lines mark the knee of the best-fit broken power law, with the shaded region showing the 16th and 84th percentiles, and the solid lines show the lines of best fit before and after the knee.}
    \label{fig:fig 5}
\end{figure}

\begin{figure}[h]
    \centering
    \includegraphics[width=0.95\linewidth]{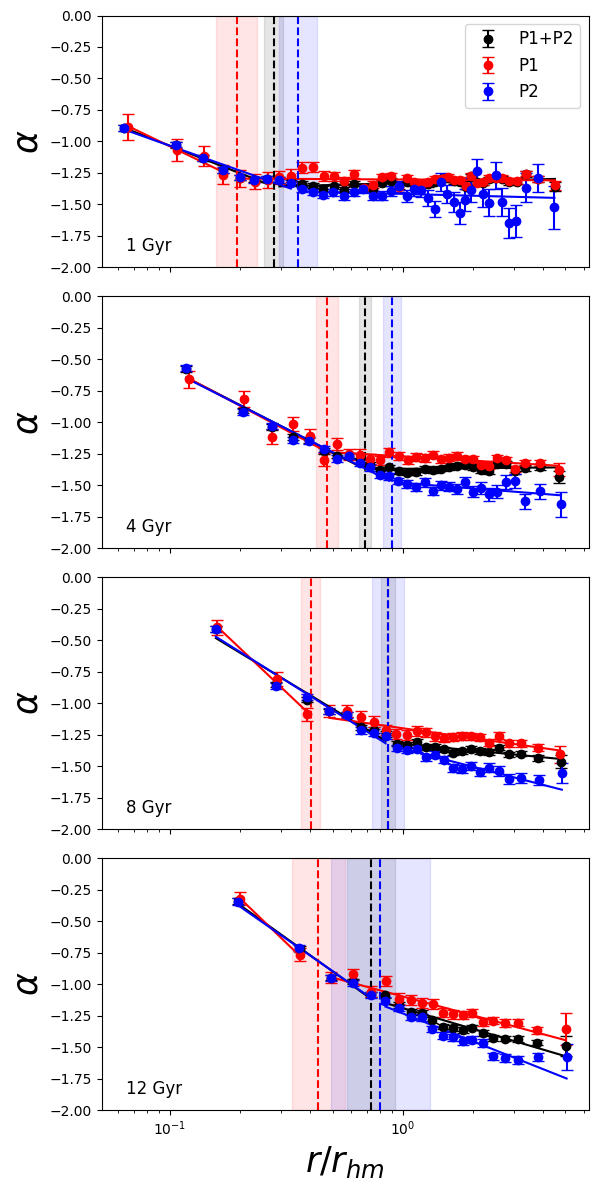}
    \caption{The slope of the stellar MF ($\alpha$) as a function of distance from the cluster center, normalized to the half-mass radius, at 1, 4, 8, and 12 Gyr snapshots within P1 (red), P2 (blue), and the total cluster (black) for the mr025c005 simulation, within the mass range $0.1$--$0.5\,M_\odot$. The vertical dashed lines mark the knee of the best-fit broken power law, with the shaded region showing the 16th and 84th percentiles, and the solid lines show the lines of best fit before and after the knee.}
    \label{fig:fig 6}
\end{figure}

\begin{figure}[h]
    \centering
    \includegraphics[width=0.95\linewidth]{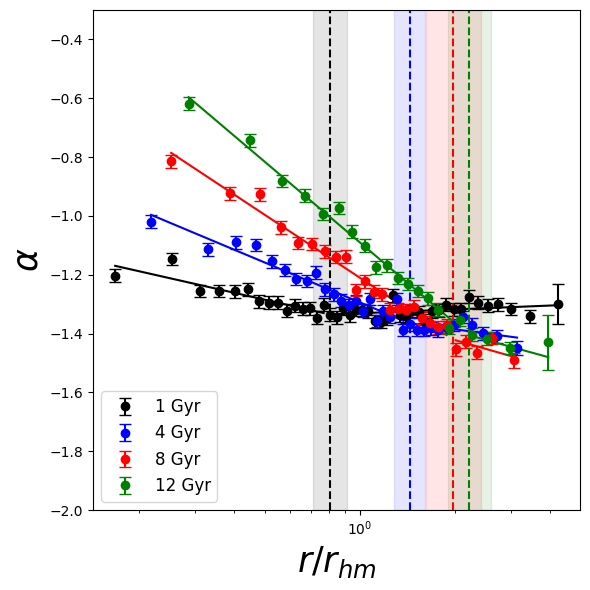}
    \caption{The slope of the stellar MF ($\alpha$) as a function of distance from the cluster center, normalized to the half-mass radius, at four different time snapshots for the SPtf simulation, within the mass range $0.1$--$0.5\,M_\odot$. The vertical dashed lines mark the knee of the best-fit broken power law, with the shaded region showing the 16th and 84th percentiles, and the solid lines show the lines of best fit before and after the knee.}
    \label{fig:fig 7}
\end{figure}

\begin{figure}[h]
    \centering
    \includegraphics[width=0.95\linewidth]{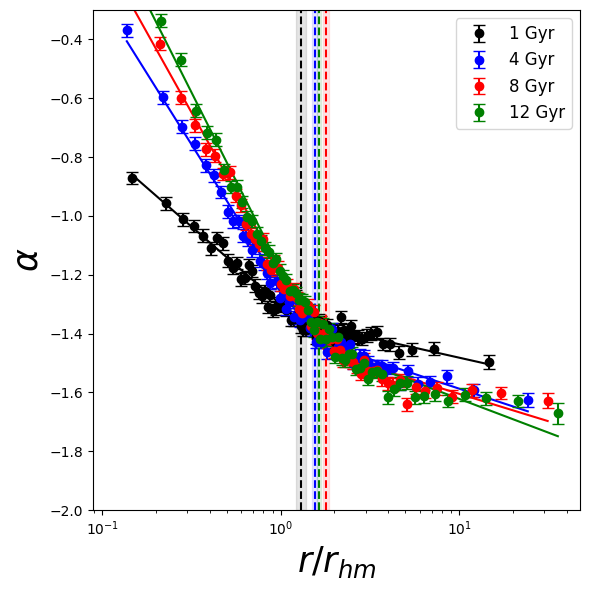}
    \caption{The slope of the stellar MF ($\alpha$) as a function of distance from the cluster center, normalized to the half-mass radius, at four different time snapshots for the SPuf simulation, within the mass range $0.1$--$0.5\,M_\odot$. The vertical dashed lines mark the knee of the best fit broken power law, with the shaded region showing the 16th and 84th percentiles, and the solid lines show the lines of best fit before and after the knee.}
    \label{fig:fig 8}
\end{figure}

\subsection{Radial variation}
\label{sec:3.1}

The evolution of the stellar MF within GCs is a complex process driven by many interconnected dynamical mechanisms, which result in the development of radial variations in the slope of the MF. Here we investigate the 3D radial profiles of the MF slope for main sequence stars in the lower mass regime ($0.1$--$0.5\,M_\odot$) of the K01 simulations, and we refer to Appendix \ref{sec:A1} for the results for different mass bins, and to Appendix \ref{sec:A2} for the results studying the radial profiles as a function of 2D projected clustercentric distances. 

In Figure~\ref{fig:fig 1} we display the radial variation of $\alpha$ at four different times. Initially, the MF slope is uniform, but over time the inner regions quickly develop a more positive slope, while the outer regions more gradually develop a more negative slope. This is expected, as two-body relaxation drives the inward segregation of more massive stars, flattening the slope toward top-heavy values, while the lower-mass stars migrate outward and dominate the outer regions, producing a bottom-heavy or more negative slope. This figure also illustrates that the slope in the inner regions of the cluster changes most dramatically from the initial conditions, as the relaxation timescales there are shortest, while the outer regions with longer relaxation timescales retain a larger fraction of their initial stellar content throughout the simulation.

This evolution is further shaped by the presence of multiple stellar populations, as highlighted in Figure~\ref{fig:fig 2}, which shows the MF slope in each radial bin computed separately within each population. In the inner regions, where the relaxation timescale is shortest, both P1 and P2 evolve similarly toward flatter, more positive slopes.

In the outer regions, however, the populations diverge, with P2 exhibiting a significantly steeper slope than P1 at all times after the initial conditions. This can be attributed to the fact that the outer regions are initially populated almost exclusively by P1 stars, where the P1 MF slope evolves gradually through standard dynamical processes. The P2 stars however, are initially sparse or completely absent in these outer regions, as reflected in the larger uncertainties on the P2 slope at large radii at 1 Gyr in Figure~\ref{fig:fig 2}, where the equal-number radial bins (matching those in Figure~\ref{fig:fig 1}) contain fewer P2 stars compared to P1. As a result, the P2 MF slope is heavily dominated by the outward migration of P2 low-mass stars initially concentrated in the cluster's inner regions (while higher-mass P2 stars preferentially remain in the central regions and segregate inward). This combination results in a significantly steeper P2 slope in the outer regions. The contributions of the two populations then produce the total radial variation displayed in Figure~\ref{fig:fig 1}, where the inner regions show a flatter slope consistent with both populations, but the outer regions show a slope intermediate between the two.

The top panel of Figure~\ref{fig:fig 3} shows the radial variation of MF slope across all K01 simulations at 12 Gyr. We highlight that the dynamical evolution of the simulations with different initial conditions results in only small differences overall, with a similar profile being found across the simulations, especially when we consider the simulations at the same dynamical age (bottom panel of Figure~\ref{fig:fig 3}). The exception is the simulation modeled in a stronger tidal field, which is characterized by a shorter relaxation time and enhanced mass loss through stellar escape, primarily of low-mass stars, flattening the MF slope and driving the cluster to a much older dynamical age. The degree of initial concentration of the P2 population has a negligible impact on the evolution of the radial profile. The simulation containing fewer P2 stars (mr01c01) does display a radial profile with central values that are not as flat as the other simulations, because the smaller P2 population results in a lower central density and slightly longer relaxation timescales in the innermost regions.

In the top panel of Figure~\ref{fig:fig 4} we show the simulations at 12 Gyr split into their multiple populations, where the same trends described above are recovered: the P2 population exhibits a more negative slope in the outer regions, while both P1 and P2 become flatter in the inner regions. Once again, the differences between the simulations are minimal, especially when considering the models at the same dynamical age (bottom panel of Figure~\ref{fig:fig 4}), showing comparable values within the radial bins independent of the concentration or fraction of P2 stars, except for the strong tidal field simulation, which displays a smaller difference between the populations due to its more advanced dynamical age, although the differences in P1 and P2 persist.

A few studies \citep[see e.g.][]{Webb+vesperini2016, Webb+2017, Cadelano+2020} have parametrized the radial gradient in the MF slope through the $\delta\alpha$ parameter, defined as the slope of the line of best fit in the space $\alpha(r)$ vs $log(R/R_{hl})$ (or $log(r/r_{hm})$ as in this work). \footnote{We note that $\delta\alpha$ is sometimes defined using the natural logarithm rather than $log_{10}$, which yields similar conclusions.} However, as shown in Figure~\ref{fig:fig 5}, the radial profile is better described by a broken power law, with a knee generally located slightly interior to the half-mass radius until late dynamical ages ($t/t_{rh}(t) >\sim10$). We fit this broken power law as two linear segments in $\alpha$ vs $log(r/r_{hm})$ that share a common value at the knee, ensuring continuity between the inner and outer fits. The knee location and value, together with the two slopes, are treated as free parameters and are determined using an MCMC approach, sampling the posterior of four parameters given the binned radial profile, with the likelihood weighted by the uncertainty in $\alpha$ in each bin. As a continuous parameter, the fitted knee location is not restricted to the discrete radial bins, and can fall between points.

This broken power law illustrates how the region interior to the half-mass radius more efficiently develops a steepening gradient over time, with the regions closer to the cluster center attaining increasingly less negative (flatter) values of $\alpha$. In contrast, the P1 stars that represent a larger fraction in the outer regions beyond the knee compared to the central regions are characterized by a MF slope that evolves away from the initial slope more slowly, as their longer relaxation timescales allow them to retain the IMF shape for a longer period of time. As a result, the outer regions are characterized by a flatter line of best fit throughout the simulation, or initially even an inverted slope such as that seen at 1 Gyr, with $\alpha$ increasing with clustercentric distance. Over time however, the two slopes gradually converge, and at very late dynamical ages a single power law would be expected to provide a sufficient description.

The presence of this knee is partially driven by the multiscale structure of the model, with the multiple stellar populations affecting both the slopes of these lines of best fit and the location of the knee itself. The relative contributions of the two stellar populations are displayed in Figure~\ref{fig:fig 6}. The total (P1+P2) value of $\alpha$ at a given radius is set primarily by the population contributing the majority of stars at that radius. Since P2 is initially more centrally concentrated than P1, P2 stars outnumber P1 stars in the inner regions, while the fraction of P1 stars increases toward the outer regions (we refer to \citet{Bruce+2026} for a discussion on the spatial mixing of the two populations in these simulations). In the inner regions, P1 and P2 develop similar radial gradients in $\alpha$, so the total profile follows this shared gradient. In the outer regions, however, P1 retains a flatter MF slope than P2, affecting the total profile. The location of the knee in the total cluster profile is therefore closely related to the radius at which the values of $\alpha$ transition from the steep inner gradient to the flatter outer P1 profile. Each population also develops its own knee, which shapes the underlying P1 and P2 profiles and further affects the location of the knee in the total profile. Interestingly, the location of the P2 knee (despite P2 being initially more centrally concentrated) is located further from the cluster center than that in P1.

To better understand the formation of this knee near the half-mass radius, we examine the radial profiles of two single-population clusters, one tidally filling (SPtf) and one tidally underfilling (SPuf), which serve as approximate dynamical proxies for P1 and P2, respectively, though we note that important differences remain between these models and individual populations within multiple-population clusters. Most notably, P2 evolves embedded within the total cluster potential (set by P1 and P2), whereas SPuf and SPtf evolve in isolation. These profiles are displayed in Figures~\ref{fig:fig 7} and \ref{fig:fig 8}. In the SPtf model (Figure~\ref{fig:fig 7}) the profile initially develops a broken profile with a knee, but soon evolves into a profile characterized by a steep radial gradient with no clear knee. This can be attributed to the fact that the local relaxation time at nearly all locations in the cluster is shorter than the age of the system. In contrast, in the SPuf model (Figure~\ref{fig:fig 8}) a distinct knee emerges, as the lower-mass stars migrating outward are more likely to be retained within the system rather than being tidally stripped or escaping through evaporation due to the underfilling nature of the cluster. The location of the knee in this case, however, lies exterior to the half-mass radius with little location variation. This location approximately corresponds to the radius at which the local relaxation time equals the age of the system, which moves outward over time as both the age of the system and the local relaxation times evolve. Stars beyond this point have relaxation timescales longer than that of the system in general, and  cannot form a steep radial gradient as efficiently.

As shown in Figure~\ref{fig:fig 8}, the presence of a knee is not strictly dependent on the presence of multiscale structure. However, for the knee to be located interior to the half-mass radius, within the models explored here, the multiscale structure associated with the presence of multiple stellar populations appears to play a key role, though this conclusion should be interpreted with caution given the limitations of the SPtf and SPuf models as proxies. 

\subsection{Global Variation}
\label{sec:3.2}

We now turn to the evolution of the global stellar MF properties, which depends on both the internal dynamics of the cluster and the strength of the tidal field. We additionally use the $\delta\alpha$ parameter, defined in Section~\ref{sec:3.1} as the slope of the best-fit single power law to the radial profile in log space ($\frac{d\alpha}{dlog(r/r_{hm})}$), to explore the link between the evolution of the slope of the global MF ($\alpha_G$) to its radial variation inside the cluster. As highlighted in Section~\ref{sec:3.1}, a single power law rarely captures the radial profile accurately due to the presence of the knee near $\sim0.6-1\,r_{hm}$, but $\delta\alpha$ nonetheless provides a useful approximation of the overall radial trend for comparison purposes. 

Figure~\ref{fig:fig 9} displays the time evolution of $\alpha_G$ and $\delta\alpha$ for the four simulations within the mass range $0.1$ -- $0.5$ $M_\odot$, and we refer to Appendix \ref{sec:A1} for the results when we consider different mass ranges. In all simulations, as the clusters evolve, the slope becomes flatter due to the preferential loss of low-mass stars, which are concentrated in the outer regions as a result of their outward migration through two-body relaxation. The evolution of $\delta\alpha$, however, shows a slight difference between the simulations: the model containing fewer P2 stars does not develop as steep a radial gradient (as expected from Section~\ref{sec:3.1}). In general, the simulations develop a radial gradient relatively quickly, after which $\delta\alpha$ plateaus at a value near $\delta\alpha\sim-0.7$--$0.8$.

Figure~\ref{fig:fig 10} shows the evolution of $\alpha_G$ and $\delta\alpha$ within each stellar population. Initially, $\alpha_G$ becomes flatter in P1 compared to P2, since the centrally concentrated P2 subsystem retains a larger fraction of its stars. As shown in the inset panel these differences are, however, very small and likely difficult to detect observationally. However, at late dynamical stages the two tracks evolve mostly in parallel as the populations spatially mix and a similar fraction of low-mass stars is lost from each. The values of $\delta\alpha$ also evolve differently between P1 and P2, with P2 developing a steeper radial gradient earlier due to the shorter relaxation timescales affecting the majority of its stars near the cluster center. In contrast, the more extended P1 takes longer to develop a comparable radial gradient. This difference in $\delta\alpha$ between the two populations then decreases at later dynamical stages, consistent with the trends discussed in Section~\ref{sec:3.1}, with the closest agreement found in the mr025c005sf model, which reaches the largest degree of spatial mixing. 

In the $\alpha_G$ vs $\delta\alpha$ panel, the two populations occupy distinct tracks throughout most of the cluster's evolution, where P2 is characterized by both a steeper global value of $\alpha_G$ and a larger radial variation in $\alpha$ between the inner and outer regions of the cluster. As the cluster evolves dynamically, the differences between the two populations decrease, although they do not fully disappear by the end of the simulations since the two populations do not reach complete spatial mixing within 12 Gyr. In \citet{Bruce+2026} we quantified the degree of spatial mixing in these simulations using the A+ parameter, which approaches zero as the populations become fully mixed. By 12 Gyr, most models reach values of A+ between $-0.06$ and $-0.08$, with the mr025c005sf model being the closest to complete mixing (A+$\approx$-0.02). Once fully mixed, the populations are expected to undergo similar dynamical evolution, and their values of $\delta\alpha$ should converge. Their values of $\alpha_G$, however, are expected to remain offset due to their different evolutionary histories prior to mixing.

\begin{figure}[h]
    \centering
    \includegraphics[width=0.95\linewidth]{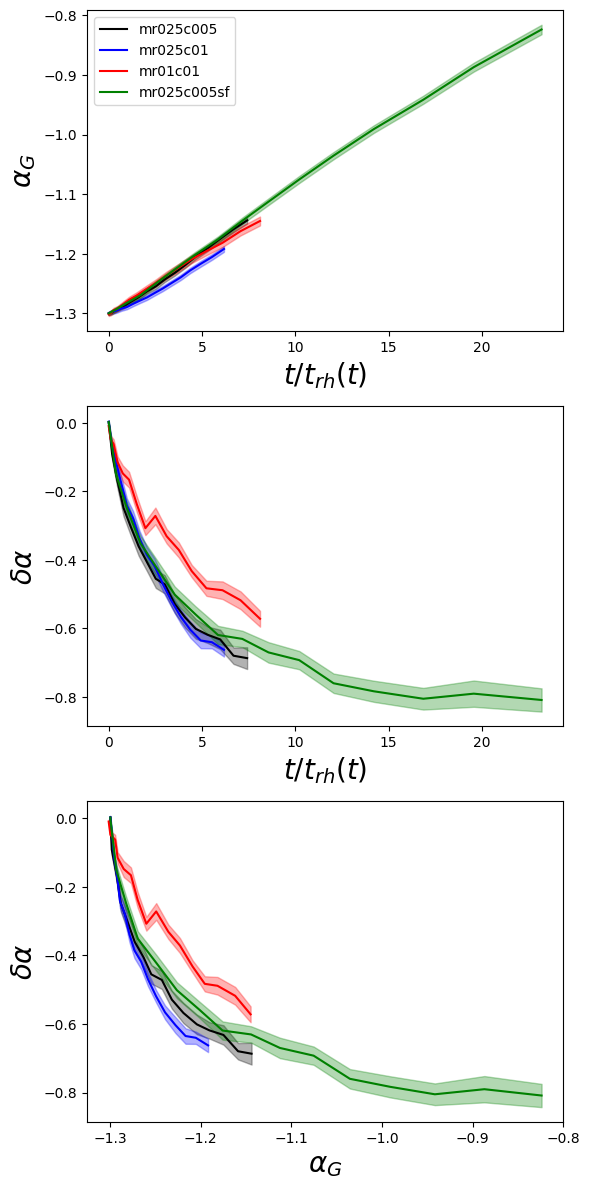}
    \caption{Time evolution of the global MF properties for four different models, within the mass range $0.1$--$0.5\,M_\odot$. (Top) The global MF slope $\alpha_G$ as a function of dynamical age $t/t_{rh}(t)$. (Middle) The time evolution of $\delta\alpha$, defined as the slope of the line of best fit to the radial profile of $\alpha$ in log space ($\frac{d\alpha}{dlog(r/r_{hm})}$), as a function of dynamical age $t/t_{rh}(t)$. (Bottom) $\delta\alpha$ as a function of the global MF slope $\alpha_G$. In all panels, the shaded regions represent the 16th and 84th percentiles of the MCMC posterior.}
    \label{fig:fig 9}
\end{figure}

\begin{figure}[h]
    \centering
    \includegraphics[width=0.95\linewidth]{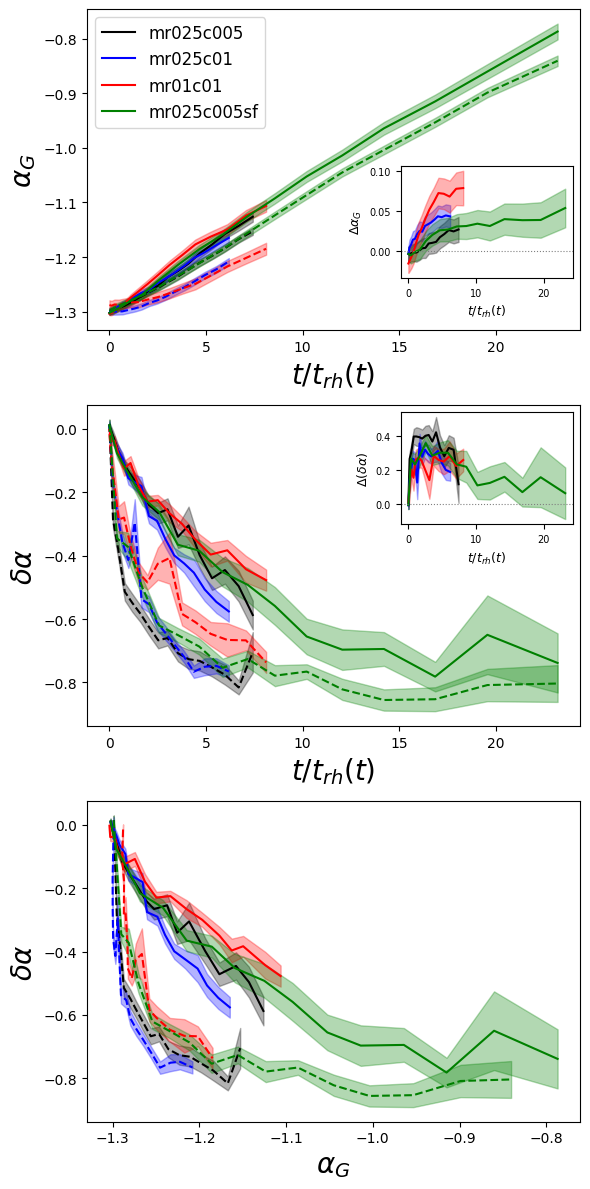}
    \caption{Time evolution of the global MF properties for four different models within P1 (solid lines) and P2 (dashed lines), within the mass range $0.1$--$0.5\,M_\odot$. (Top) The global MF slope $\alpha_G$ as a function of dynamical age $t/t_{rh}(t)$. (Middle) The time evolution of $\delta\alpha$, defined as the slope of the line of best fit to the radial profile of $\alpha$ in log space ($\frac{d\alpha}{dlog(r/r_{hm})}$), as a function of dynamical age $t/t_{rh}(t)$. (Bottom) $\delta\alpha$ as a function of the global MF slope $\alpha_G$. In all panels, the shaded regions represent the 16th and 84th percentiles of the MCMC posterior.}
    \label{fig:fig 10}
\end{figure}

\subsection{Comparison between local and global MF slopes}
\label{sec:3.3}

A parameter of particular significance for observational studies is the location at which the local value of $\alpha$ is equal to the global value ($r_{Glob})$, which can be seen in Figure~\ref{fig:fig 1} for one of the models at four different time steps, and in Figure~\ref{fig:fig 3} for the four models at 12 Gyr. The locations $r_{Glob}$ and $r_{IMF}$ are defined as the radii at which the radial profile of $\alpha$ crosses the global slope and the IMF slope, respectively. These crossing radii are identified by linearly interpolating between the first pair of adjacent bins where the profile crosses the target value. In cases where no crossing is found, the snapshot is excluded from analysis. In cases where multiple crossings exist due to noise, we report the innermost crossing. As the cluster evolves, this location moves outward, as both the global value of $\alpha$ and the radial profile evolve. This outward migration is well described by a logarithmic dependence on dynamical age, with $r_{Glob}/r_{hm}$ increasing linearly as a function of $log(t/t_{rh}(t))$ as demonstrated in Figure~\ref{fig:fig 11}. 

One can similarly examine this location within each stellar population, where the local $\alpha$ within a population is equal to the global value of that population. This is displayed for one simulation at four different times in Figure~\ref{fig:fig 2}, and for four different models at 12 Gyr in Figure~\ref{fig:fig 4}. Interestingly, as the simulations evolve, this location moves outward slightly more rapidly within P1 than within P2, despite P2 being initially more centrally concentrated. This behavior can be partly attributed to the preferential loss of low-mass P1 stars through tidal stripping and evaporation in the outer regions, which flattens the global value of $\alpha$ for P1, whereas the majority of P2 stars are retained and the global P2 value of $\alpha$ remains relatively constant. Figure~\ref{fig:fig 12} shows the outward migration of this location continues to progress linearly with $log(t/t_{rh}(t))$, although P1 shows a slightly steeper trend.

A similarly interesting location is the radius at which the local value of $\alpha$ is equal to the IMF value ($r_{IMF}$). This point evolves more dramatically than $r_{Glob}$ due to the additional complexities associated with the global slope of the MF evolving away from the IMF. This location can be located over time in Figure~\ref{fig:fig 1}, and across the four models at 12 Gyr in Figure~\ref{fig:fig 3}. Unlike $r_{glob}$, $r_{IMF}$ evolves outward almost linearly with dynamical age as stars are lost from the system. Figure~\ref{fig:fig 13} shows the steady outward migration of $r_{IMF}/r_{hm}$, which evolves consistently across the different K01 simulations with little variation. 

As seen previously in Figures~\ref{fig:fig 2} and \ref{fig:fig 4}, and further displayed in Figure~\ref{fig:fig 14}, the evolution of $r_{IMF}$ differs significantly between the multiple populations, largely due to both the enhanced rate of stellar escape in P1 and the shorter relaxation timescales within P2. As a result, the location of this point in P1 moves outward rapidly, while the location within P2 evolves at a significantly slower rate, making it a more stable candidate for observational study. Both the outward migration of $r_{IMF}$ within P1 and P2 however, continue to be well described as a linear function of dynamical age. We refer to Appendix \ref{sec:A1} for the results within different mass ranges, and to Appendix \ref{sec:A2} for the results using 2D projected clustercentric distances.

\begin{figure}[h]
    \centering
    \includegraphics[width=0.95\linewidth]{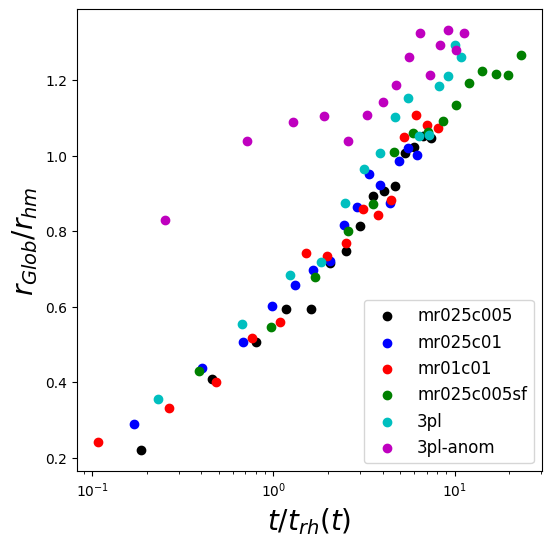}
    \caption{Time evolution of the location at which the local value of $\alpha$ equals the global value (measured at time $t$), for six different simulations, as a function of dynamical age ($t/t_{rh}(t)$), within the mass range $0.1$--$0.5\,M_\odot$ (K01) or $0.1$--$0.4\,M_\odot$ (3pl and 3pl-anom).}
    \label{fig:fig 11}
\end{figure}

\begin{figure}[h]
    \centering
    \includegraphics[width=0.95\linewidth]{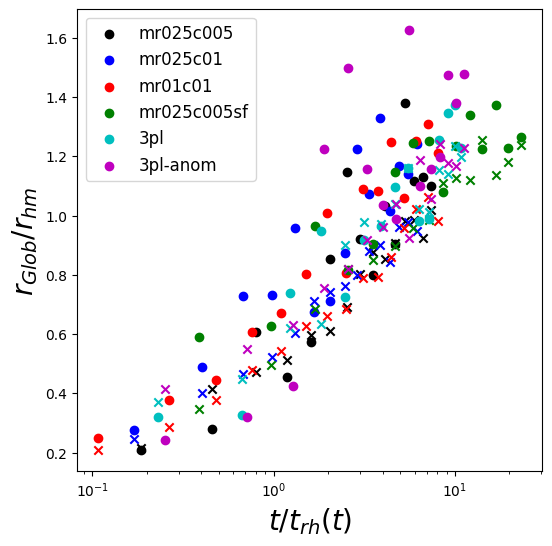}
    \caption{Time evolution of the location at which the local value of $\alpha$ equals the global value {(measured at time $t$)} within each population, for six different simulations, as a function of dynamical age ($t/t_{rh}(t)$), for P1 (circles) and P2 (crosses), within the mass range $0.1$--$0.5\,M_\odot$ (K01) or $0.1$--$0.4\,M_\odot$ (3pl and 3pl-anom).}
    \label{fig:fig 12}
\end{figure}

\begin{figure}[h]
    \centering
    \includegraphics[width=0.95\linewidth]{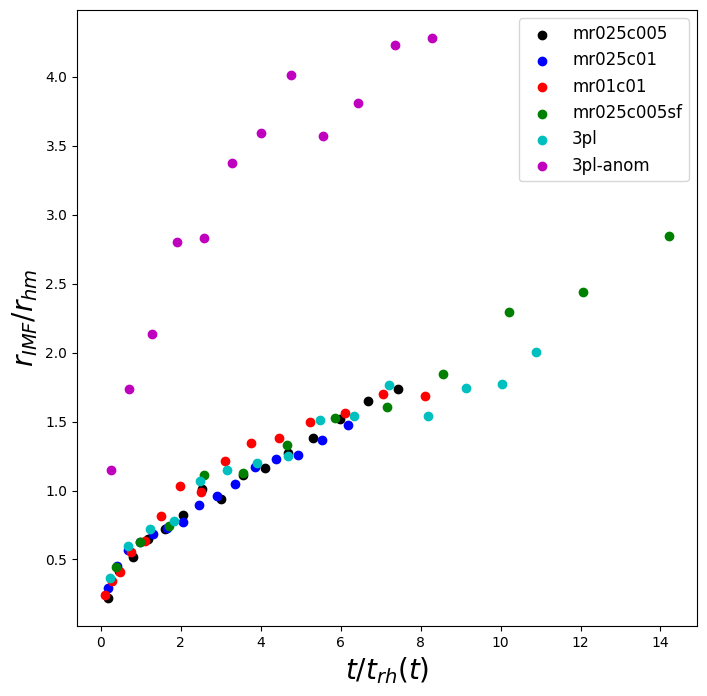}
    \caption{Time evolution of the location where the local value of $\alpha$ equals the initial value, for six different simulations as a function of dynamical age ($t/t_{rh}(t)$), within the mass range $0.1$--$0.5\,M_\odot$ (K01) or $0.1$--$0.4\,M_\odot$ (3pl and 3pl-anom).}
    \label{fig:fig 13}
\end{figure}

\begin{figure}[h]
    \centering
    \includegraphics[width=0.95\linewidth]{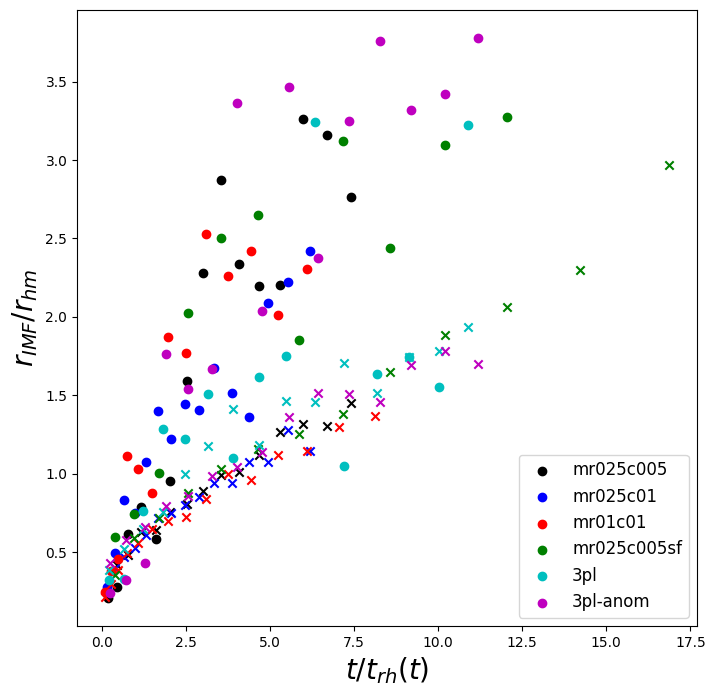}
    \caption{Time evolution of the location at which the local value of $\alpha$ equals the IMF value within six different simulations as a function of dynamical age ($t/t_{rh}(t)$) for P1 (circles) and P2 (crosses), within the mass range $0.1$--$0.5\,M_\odot$ (K01) or $0.1$--$0.4\,M_\odot$ (3pl and 3pl-anom).}
    \label{fig:fig 14}
\end{figure}

\subsection{Bottom-light IMF and different IMFs for P1 and P2}
\label{sec:3.4}

\begin{figure}[h]
    \centering
    \includegraphics[width=0.95\linewidth]{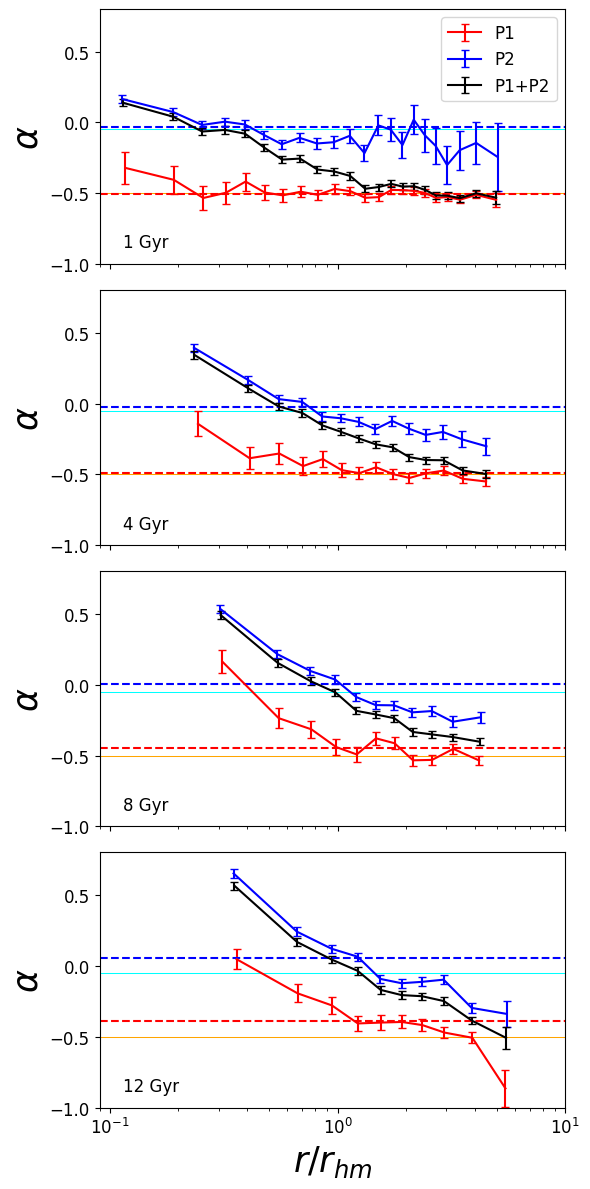}
    \caption{The slope of the stellar MF ($\alpha$) as a function of distance from the cluster center, normalized to the half-mass radius, at 1, 4, 8, and 12 Gyr snapshots within P1 (red) and P2 (blue) for the 3pl-anom simulation, within the mass range $0.1$--$0.4\,M_\odot$. The horizontal lines display the current global value of $\alpha$ within each population. The horizontal cyan (P2) and orange (P1) lines display the initial values of $\alpha$ within each population.}
    \label{fig:fig 15}
\end{figure}

\begin{figure}[h]
    \centering
    \includegraphics[width=0.95\linewidth]{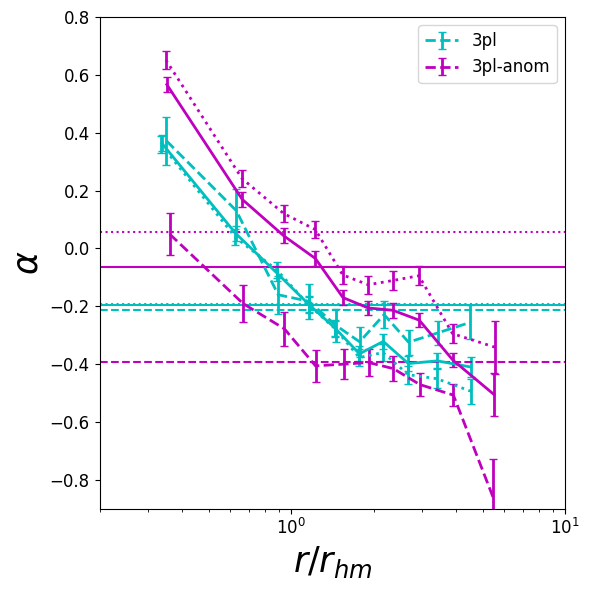}
    \caption{The slope of the stellar MF ($\alpha$) as a function of distance from the cluster center, normalized to the half-mass radius, at 12 Gyr for the 3pl (cyan) and 3pl-anom (magenta) simulations within P1 (dashed lines), P2 (dotted lines), and P1 + P2 (solid lines), within the mass range $0.1$--$0.4\,M_\odot$. The horizontal lines display the current global value of $\alpha$ within each population.}
    \label{fig:fig 16}
\end{figure}

\begin{figure}[h]
    \centering
    \includegraphics[width=0.95\linewidth]{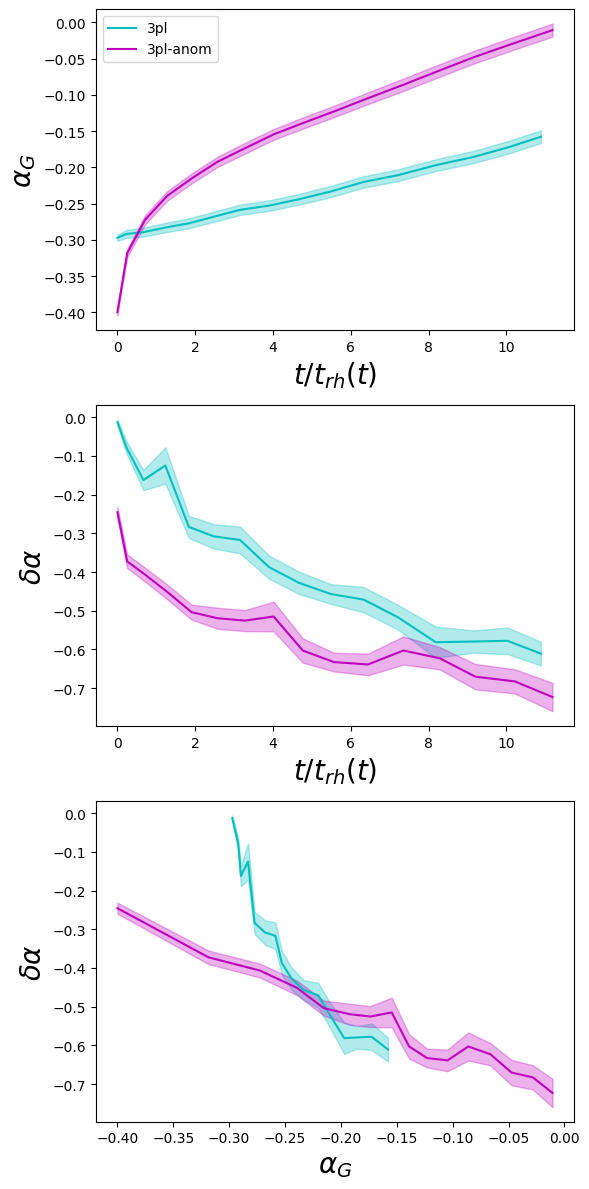}
    \caption{Time evolution of the global MF properties for the 3pl and 3pl-anom simulations, within the mass range $0.1$--$0.4\,M_\odot$. (Top) The global MF slope $\alpha_G$ as a function of dynamical age $t/t_{rh}(t)$. (Middle) The time evolution of $\delta\alpha$, defined as the slope of the line of best fit to the radial profile of $\alpha$ in log space ($\frac{d\alpha}{dlog(r/r_{hm})}$), as a function of dynamical age $t/t_{rh}(t)$. (Bottom) $\delta\alpha$ as a function of the global MF slope $\alpha_G$. In all panels, the shaded regions represent the 16th and 84th percentiles of the MCMC posterior.}
    \label{fig:fig 17}
\end{figure}

\begin{figure}[h!]
    \centering
    \includegraphics[width=0.95\linewidth]{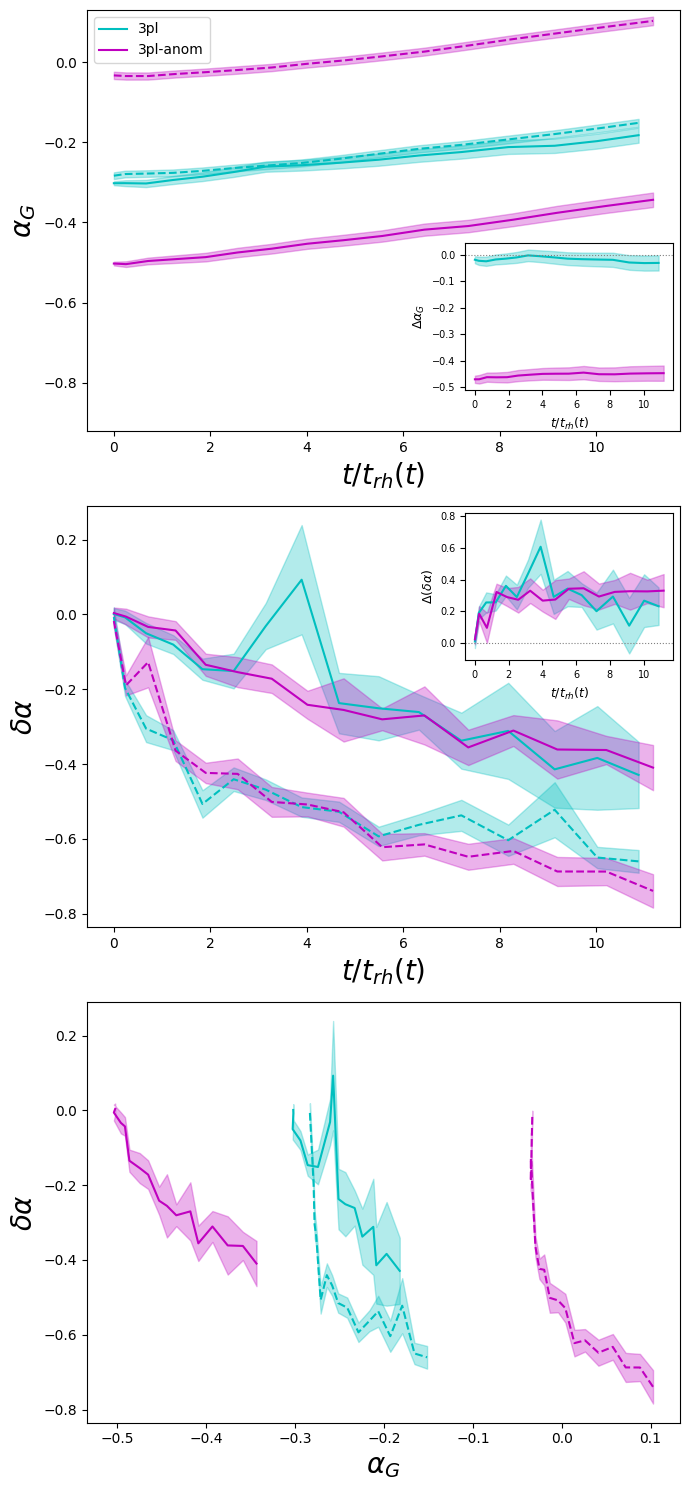}
    \caption{Time evolution of the global MF properties for the 3pl and 3pl-anom simulations within P1 (solid lines) and P2 (dashed lines), within the mass range $0.1$--$0.4\,M_\odot$. (Top) The global MF slope $\alpha_G$ as a function of dynamical age $t/t_{rh}(t)$. (Middle) The time evolution of $\delta\alpha$, defined as the slope of the line of best fit to the radial profile of $\alpha$ in log space ($\frac{d\alpha}{dlog(r/r_{hm})}$), as a function of dynamical age $t/t_{rh}(t)$. (Bottom) $\delta\alpha$ as a function of the global MF slope $\alpha_G$. In all panels, the shaded regions represent the 16th and 84th percentiles of the MCMC posterior.}
    \label{fig:fig 18}
\end{figure}

In this section we examine the dynamical evolution of a bottom-light three-parameter IMF (3pl), as well as the 3pl-anom model, in which the P2 IMF is characterized by a flatter slope in the lower mass range ($0.1$ -- $0.4\,M_\odot$). Unlike the K01 and 3pl models, where the P1 and P2 share the same IMF and the initial (at t=0 Gyr) combined MF slope is therefore spatially uniform, the total (P1+P2) profile in the 3pl-anom model already shows a radial gradient in $\alpha$ at t=0. This is the result of the differing MP IMFs combined with the different initial spatial concentrations (with a more centrally concentrated P2). Any additional radial variations seen at later times then develop from this radial gradient rather than a flat starting point. 

The overall evolution of the mass function properties of a cluster starting with the 3pl IMF is qualitatively similar to the K01 models described above in Sections~\ref{sec:3.1}, \ref{sec:3.2} and \ref{sec:3.3}: the radial profile develops a broken power law with a knee near the half-mass radius, a flatter MF slope in the inner regions, and a steeper MF slope in the outer regions. Globally, the MF slope flattens over time as low-mass stars are preferentially lost, but due to the flatter IMF the global MF slope remains slightly flatter than the K01 simulations, and it does not develop as steep of a radial gradient. The location of $r_{Glob}$ and $r_{IMF}$ also closely follow the trends in the K01 simulations as displayed in Figures~\ref{fig:fig 11}--\ref{fig:fig 14}. For the 3pl-anom model, the values of $r_{Glob}$ and $r_{IMF}$ are computed using the combined (P1+P2) MF slope at each snapshot or in the initial snapshot (with $\alpha$ equal to $\sim-0.4$), respectively. The trends in the evolution of the MF properties for the 3pl-anom model are also qualitatively similar when the P1 and P2 are combined together, but the evolution of the MF, calculated separately for each population, varies significantly. 

In Figure~\ref{fig:fig 15} we display the value of $\alpha$ as a function of distance from the cluster center at four different times for the 3pl-anom model. As in the simulations described earlier, both populations evolve such that the central regions become flatter as high-mass stars segregate inward, while the outer regions become steeper than the global value as lower-mass stars migrate outward. However, because P2 begins with a flatter slope in the lower mass range, it now maintains the flatter value of $\alpha$ at all radii throughout the simulation, even in the innermost regions where previously the MF slopes of P1 and P2 were consistent (see Section~\ref{sec:3.1}). This is distinctly different from the case in which both populations share the same IMF, where P2 was instead characterized by a steeper slope in the outer regions and a similar slope in the central regions. In this model, P2 exhibits a uniquely flatter value of $\alpha$ relative to P1, even resulting in a significantly inverted slope in the inner regions (indicating a larger fraction of higher-mass stars relative to lower-mass).

In Figure~\ref{fig:fig 16} we compare the 3pl and 3pl-anom simulations to illustrate how, despite producing a similar total (P1+P2) slope in the radial bins, the two models can exhibit opposite trends between P1 and P2. The 3pl simulation behaves as expected from the standard models, with the P2 MF slope being steeper than P1 in the outer regions, whereas the 3pl-anom simulation shows a flatter P2 MF slope relative to P1. Additionally, in the 3pl-anom simulation the inner regions display differing MF slopes between the two populations, which distinguishes it from the standard 3pl model, where the inner regions show similar values consistent with the identical IMFs in both populations. Although we consider a slightly different mass range and our models are not intended to reproduce 47 Tuc, the 3pl-anom model is qualitatively consistent with \citet{Scalco+2026}, who find a flatter MF slope in P2 than in P1 in the low-mass regime.

Figure~\ref{fig:fig 17} shows the time evolution of the global MF slope ($\alpha_G$) and the radial gradient parameter ($\delta\alpha$) for the 3pl and 3pl-anom simulations. Both models evolve qualitatively similar to the K01 simulations, with $\alpha_G$ flattening over time, but the 3pl-anom model begins with a steeper $\alpha_G$ and develops a flatter value by the end of the simulation. In addition to reaching a flatter $\alpha_G$, the 3pl-anom simulation also begins with a non-zero value of $\delta\alpha$ due to the differing IMFs in P1 and P2 (see discussion at the beginning of this section); additionally the 3pl-anom model develops a steeper radial gradient in MF slope, which is expected since the slope in the inner regions is initially dominated by P2 which began with a flatter IMF compared to P1, which represent a larger fraction in the outer regions. This is also reflected in the total (P1+P2) profile shown in Figure~\ref{fig:fig 16} (see also Fig~\ref{fig:fig 15}). Since P2 is more centrally concentrated than P1, its relative number fraction is largest near the cluster center and decreases with clustercentric radius, so the combined-population value of $\alpha$ is closer to the P2 value in the central regions and progressively approaches the P1 value in the outer regions. Figure~\ref{fig:fig 18} highlights this by showing the global evolution of the properties within each population. In both models, P2 develops a steeper radial gradient than P1 as discussed in Section~\ref{sec:3.2}, and while $\alpha_G$ becomes flatter over time, the P1 and P2 slopes vary significantly in the 3pl-anom model due to the differences between the P1 and the P2 IMF, while the 3pl simulation shows more similar values of $\alpha_G$.

We conducted a preliminary investigation into the effect of assigning P1 and P2 different maximum stellar masses in the IMF, adopting a maximum P2 stellar mass of $20\,M_\odot$ while keeping the P1 maximum at $100\,M_\odot$ (all simulations presented in this paper share the same maximum stellar mass of $100\,M_\odot$ for both populations). This change has only a minor effect on our results: the lower maximum P2 stellar mass slightly slows mass segregation in the central regions, reducing the flattening of $\alpha$ at radial bins within the half-mass radius in both populations, though the effect is small (<0.1 in $\alpha$) and the qualitative behavior of the radial and global trends discussed above remains unchanged. A more thorough investigation of the effects of population-dependent maximum stellar masses, including a wider range of values and its interplay with the different P1 and P2 IMF slopes explored above, is left to future work.

\section{Conclusions}
\label{sec:4}

In this paper we have investigated the dynamical evolution of the stellar MF within multiple-population globular clusters using a suite of Monte Carlo simulations run with the MOCCA code. We presented the evolution of the radial profiles of the stellar MF, characterized several interesting locations and properties within these profiles, and examined the evolution of global MF properties.

The presence of an initially more centrally concentrated second population (P2), formed embedded in a more extended first population (P1), has distinct implications for the dynamical evolution of a star cluster and its stellar MF. The multiscale structure results in distinct dynamical environments occupied by the stars of each population, leading to differing evolutionary pathways and signatures imprinted on the radial profiles. As a result, the presence of multiple stellar populations must be considered when interpreting the total stellar MF of a cluster.

As the cluster evolves, two-body relaxation drives the formation of a radial gradient in the MF slope, where the inner regions become flatter due to the inward segregation of higher-mass stars, while the outer regions become steeper as lower-mass stars migrate outward (see Figure~\ref{fig:fig 1}). This total profile arises from the combination of P1, which maintains a flatter slope in the outer regions where stars reside in environments with longer relaxation timescales, and P2, which develops a steeper slope in the outer regions as lower-mass stars preferentially migrate outward into regions initially devoid of P2 stars (see Figure~\ref{fig:fig 2}). This profile appears qualitatively similar across models with different initial P2 concentrations and fractions, shifting only toward flatter values globally at late dynamical stages (see Figures~\ref{fig:fig 3} and \ref{fig:fig 4}).

The radial variations of the slope of the MF with the clustercentric distance are well described by a broken power law with a knee located slightly interior to the half-mass radius (see Figure~\ref{fig:fig 5}), which is partially driven by the relative contributions of the two stellar populations (see Figure~\ref{fig:fig 6}). This multiscale structure is important, as radial profiles with knee locations that remain interior to the half-mass radius until late dynamical stages are difficult to reconcile with single-population clusters (see Figures~\ref{fig:fig 7} and \ref{fig:fig 8}).

Globally, the slope of the stellar MF flattens over time as a result of the preferential loss of low-mass stars, while the radial gradient of the MF slope steepens (see Figure~\ref{fig:fig 9}). The evolution of these global properties depends on the contributions from the two stellar populations, which evolve differently over time, although the $\delta\alpha$ values converge at late dynamical stages (see Figure~\ref{fig:fig 10}).

Many previous studies have sought to identify the radial distance from a cluster's center where, at any given time, the local slope of the stellar MF matches its global value. We find that this location migrates outward linearly with $log(t/t_{rh}(t))$ as the cluster evolves with relatively small scatter between our models (see Figure~\ref{fig:fig 11}), with the P1 location migrating outward at a faster rate compared to P2 (see Figure~\ref{fig:fig 12}).

An additional location, even more valuable for our understanding of cluster formation and evolution, is the radius at which the local slope of the stellar MF at a given time equals the slope of the IMF. This point evolves outward linearly with dynamical age as the global slope is continuously evolving away from the IMF when low-mass stars are preferentially lost from the cluster (see Figure~\ref{fig:fig 13}). Furthermore, while P1 stars are lost at a higher rate, causing this point to migrate outward more rapidly in P1, the corresponding location in P2 shows a much lower degree of outward migration (see Figure~\ref{fig:fig 14}).

Motivated by recent studies suggesting that a bottom-light IMF may be better suited to massive Galactic globular clusters, we ran an additional simulation adopting a three-parameter bottom-light IMF, which evolved qualitatively similar to the \citet{Kroupa2001} models, but differed slightly in the global MF slope and degree of radial gradient formed. Additionally, motivated by a recent observational study by \citet{Scalco+2026} hinting at possible differences between the P1 and the P2 IMF, we run a bottom-light three-parameter model featuring, in the lower mass range, a nearly flat P2 IMF slope and a steeper P1 IMF. The evolution of the radial profile of the slope of the MF in this model shows that, as a cluster evolves, the P2 MF slope remains flatter than the P1 MF slope at all clustercentric distances thus revealing at all times the fingerprints of the differences between the P1 and P2 IMF (see Figure~\ref{fig:fig 15} and Figure~\ref{fig:fig 16}). We investigate the effect of this bottom-light model on the variation in the global properties of the total (P1 + P2) MF (see Figure~\ref{fig:fig 17}), as well as within P1 and P2 where the global properties differ significantly (see Figure~\ref{fig:fig 18}).

Our results emphasize the importance of the multiscale structure introduced by multiple stellar populations for the evolution of both the radial profiles and the global values of the stellar MF slope. We further highlight that bottom-light IMF models evolve qualitatively similar to standard IMF models, but can produce distinct radial and global values. The results presented in this paper provide a theoretical framework to interpret the results of existing and future observational studies of the P1 and P2 MF and disentangle the effects of the clusters' dynamical evolution on the MF from those associated to possible differences between the P1 and P2 IMF.

\begin{acknowledgements}

EV acknowledges support from NSF grant AST-2009193. This research was supported in part by Lilly Endowment, Inc., through its support for the Indiana University Pervasive Technology Institute. AA acknowledges that this research was funded in part by National Science Centre (NCN), Poland, grant No. 2024/55/D/ST9/02585. For the purpose of Open Access, the authors have applied for a CC-BY public copyright license to any author Accepted Manuscript (AAM) version arising from this submission.

\end{acknowledgements}

\bibliographystyle{aa}
\bibliography{ref}

\begin{appendix}
\section{Different mass ranges}
\label{sec:A1}

In this section we present the results in two different mass ranges, $0.5$--$0.8\,M_\odot$ and $0.3$--$0.8\,M_\odot$, to demonstrate that the trends described in the main text are robust across different stellar mass regimes. We show characteristic plots for the global and local values of the MF slopes; where figures are omitted, the trends are qualitatively consistent with those presented in the main text. We follow the same process as described in Sec~\ref{sec:2}, where we restrict our analysis to single main sequence stars within the specified mass range.

Figures~\ref{fig:fig M1} and \ref{fig:fig M2} show the radial variation in the MF slope at 12 Gyr for the total cluster within the $0.5$--$0.8\,M_\odot$ and $0.3$--$0.8\,M_\odot$ mass ranges, respectively, while Figures~\ref{fig:fig M3} and \ref{fig:fig M4} show the corresponding values within each stellar population. Figures~\ref{fig:fig M5}, \ref{fig:fig M6}, \ref{fig:fig M5MP}, and \ref{fig:fig M6MP} show the time evolution of the global MF properties for the total cluster and within each population for these mass ranges, respectively. Figures~\ref{fig:fig M7}, \ref{fig:fig M8}, \ref{fig:fig M9}, \ref{fig:fig M10} show the same for the 3pl and 3pl-anom simulations. Figures~\ref{fig:fig M11} and \ref{fig:fig M12} show the radial variation for the 3pl and 3pl-anom simulations split by population. Finally, Figures~\ref{fig:fig M13}--\ref{fig:fig M16} show the time evolution of the radial characteristic locations ($r_{Glob}$ and $r_{IMF}$) within the $0.3$ -- $0.8\,M_\odot$ mass range.

\begin{figure}[h]
    \centering
    \includegraphics[width=0.95\linewidth]{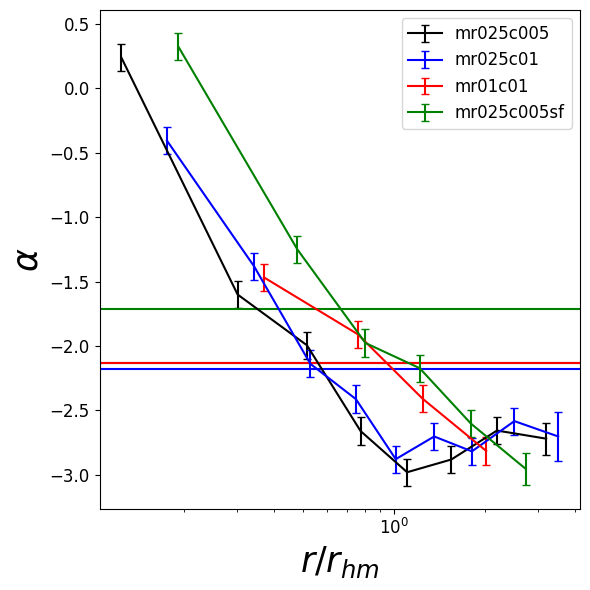}
    \caption{Same as the top panel of Figure~\ref{fig:fig 3}, but within the mass range $0.5$--$0.8\,M_\odot$.}
    \label{fig:fig M1}
\end{figure}

\begin{figure}[h]
    \centering
    \includegraphics[width=0.95\linewidth]{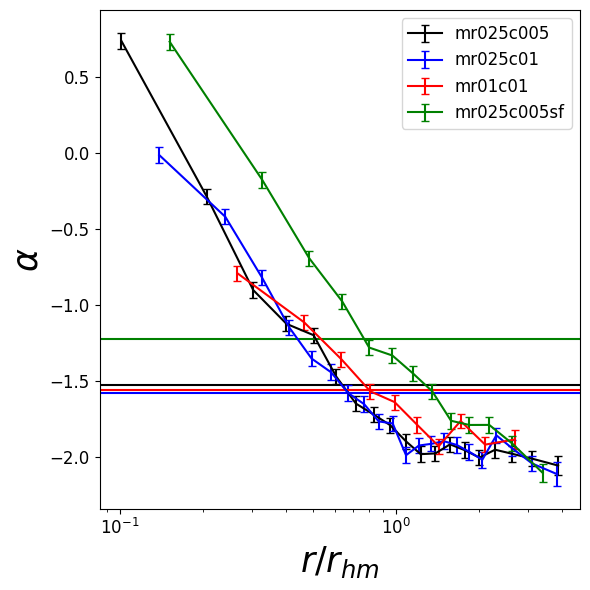}
    \caption{Same as the top panel of Figure~\ref{fig:fig 3}, but within the mass range $0.3$--$0.8\,M_\odot$.}
    \label{fig:fig M2}
\end{figure}

\begin{figure}[h]
    \centering
    \includegraphics[width=0.95\linewidth]{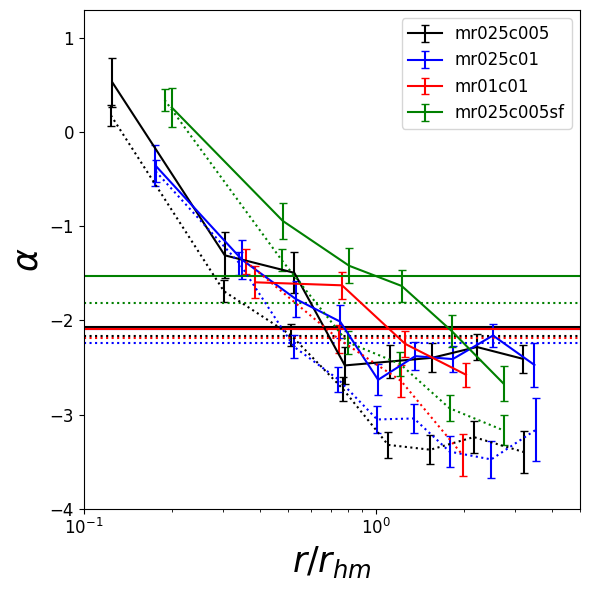}
    \caption{Same as the top panel of Figure~\ref{fig:fig 4}, but within the mass range $0.5$--$0.8\,M_\odot$.}
    \label{fig:fig M3}
\end{figure}

\begin{figure}[h]
    \centering
    \includegraphics[width=0.95\linewidth]{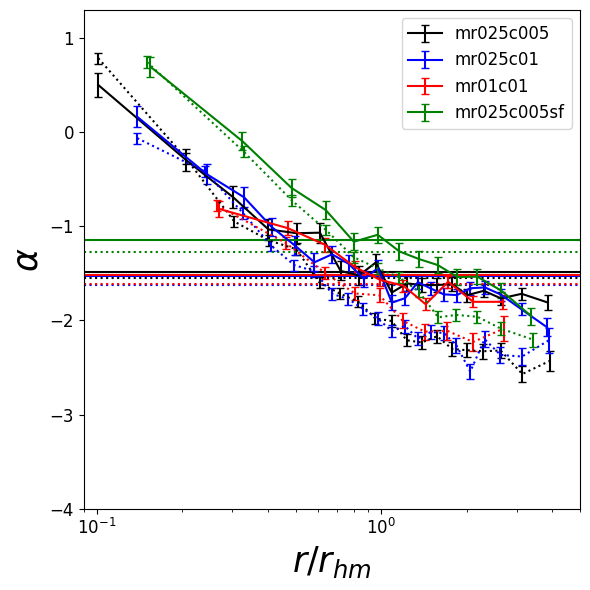}
    \caption{Same as the top panel of Figure~\ref{fig:fig 4}, but within the mass range $0.3$--$0.8\,M_\odot$.}
    \label{fig:fig M4}
\end{figure}

\begin{figure}[h]
    \centering
    \includegraphics[width=0.95\linewidth]{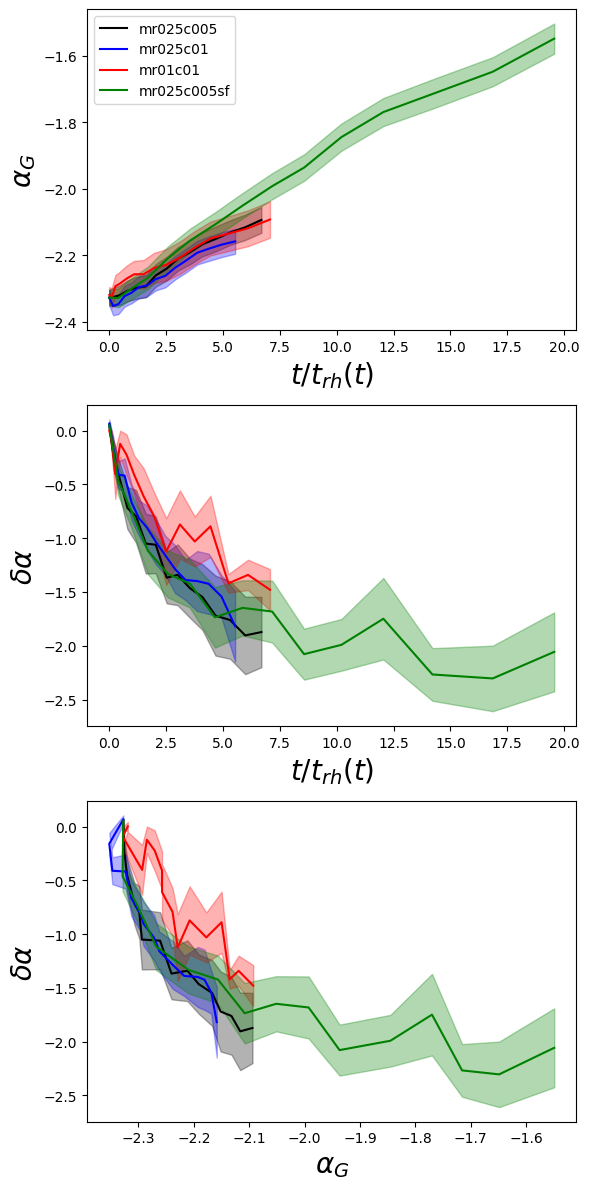}
    \caption{Same as Figure~\ref{fig:fig 9}, but within the mass range $0.5$--$0.8\,M_\odot$.}
    \label{fig:fig M5}
\end{figure}

\begin{figure}[h]
    \centering
    \includegraphics[width=0.95\linewidth]{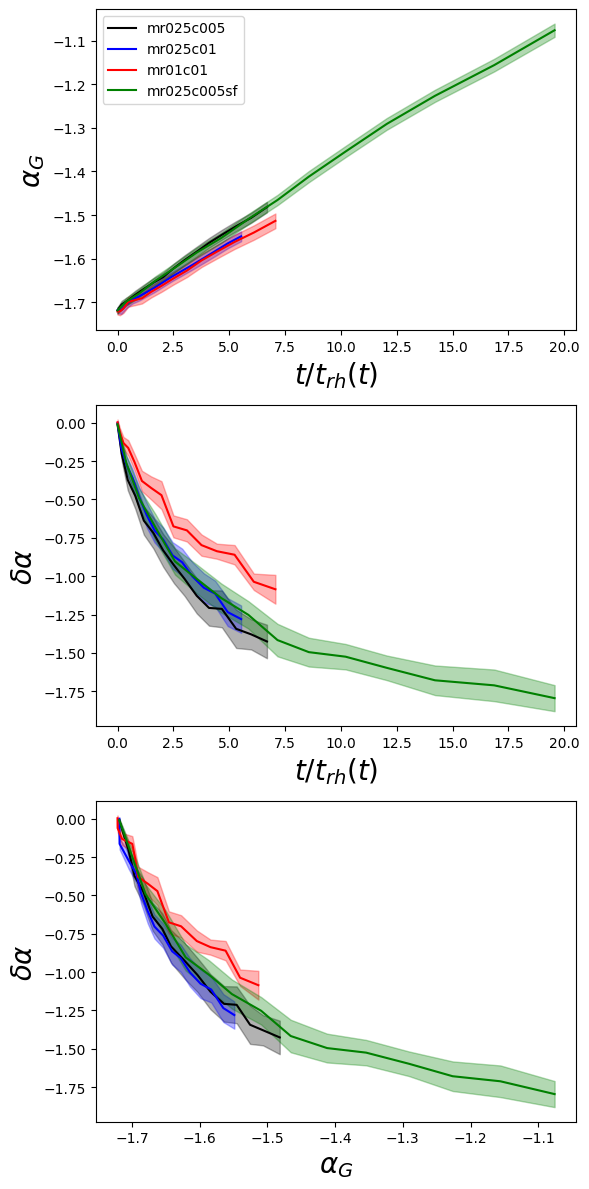}
    \caption{Same as Figure~\ref{fig:fig 9}, but within the mass range $0.3$--$0.8\,M_\odot$.}
    \label{fig:fig M6}
\end{figure}

\begin{figure}[h]
    \centering
    \includegraphics[width=0.95\linewidth]{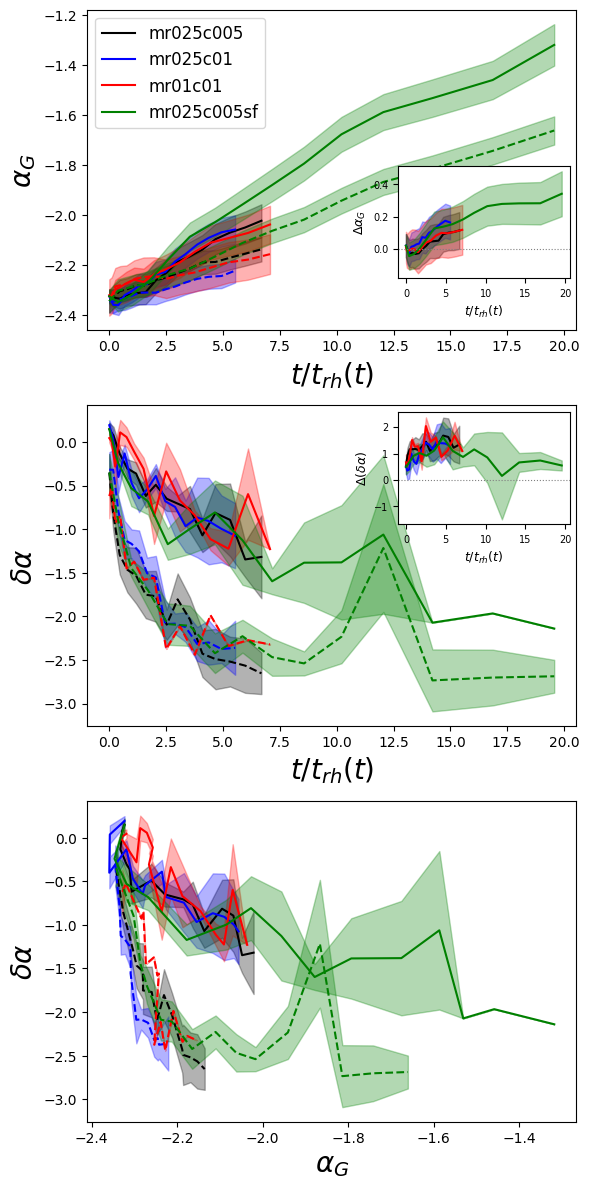}
    \caption{Same as Figure~\ref{fig:fig 10}, but within the mass range $0.5$--$0.8\,M_\odot$.}
    \label{fig:fig M5MP}
\end{figure}

\begin{figure}[h]
    \centering
    \includegraphics[width=0.95\linewidth]{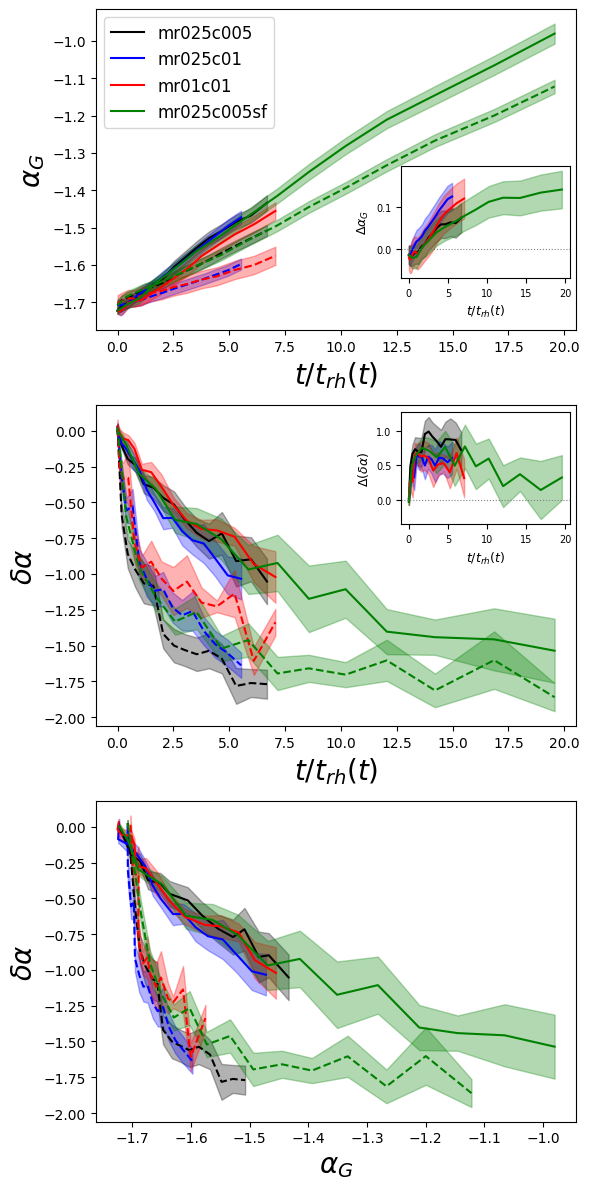}
    \caption{Same as Figure~\ref{fig:fig 10}, but within the mass range $0.3$--$0.8\,M_\odot$.}
    \label{fig:fig M6MP}
\end{figure}

\begin{figure}[h]
    \centering
    \includegraphics[width=0.95\linewidth]{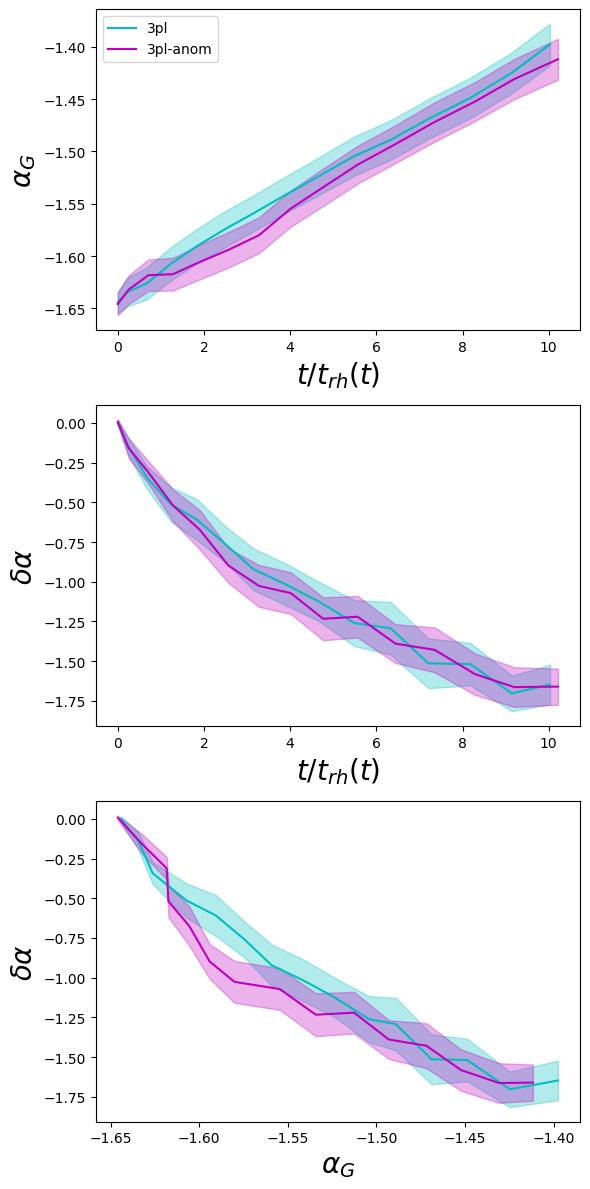}
    \caption{Same as Figure~\ref{fig:fig 17}, but within the mass range $0.5$--$0.8\,M_\odot$.}
    \label{fig:fig M7}
\end{figure}

\begin{figure}[h]
    \centering
    \includegraphics[width=0.95\linewidth]{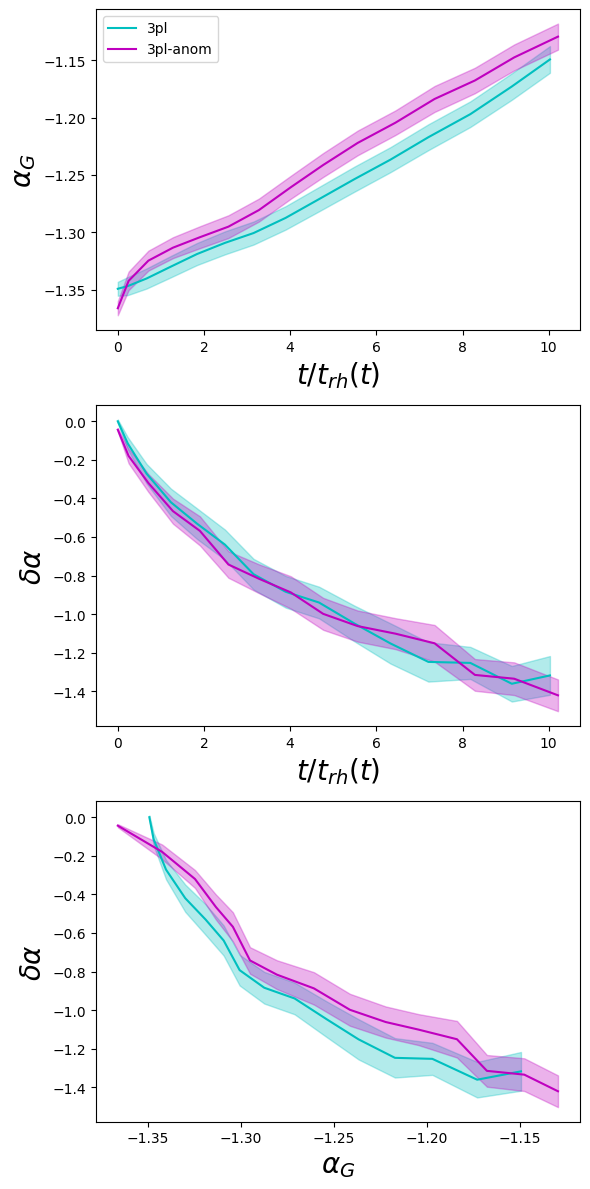}
    \caption{Same as Figure~\ref{fig:fig 17}, but within the mass range $0.3$--$0.8\,M_\odot$.}
    \label{fig:fig M8}
\end{figure}

\begin{figure}[h]
    \centering
    \includegraphics[width=0.95\linewidth]{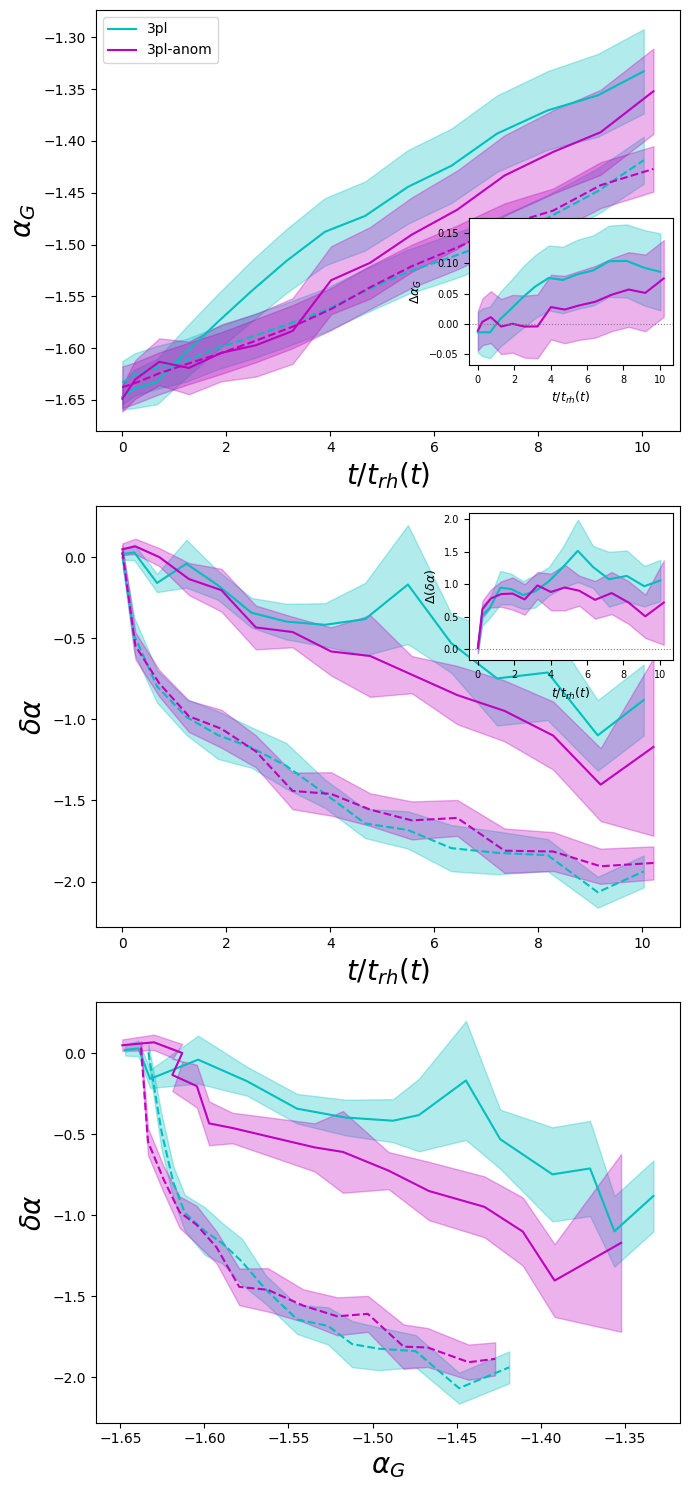}
    \caption{Same as Figure~\ref{fig:fig 18}, but within the mass range $0.4$--$0.8\,M_\odot$.}
    \label{fig:fig M9}
\end{figure}

\begin{figure}[h]
    \centering
    \includegraphics[width=0.95\linewidth]{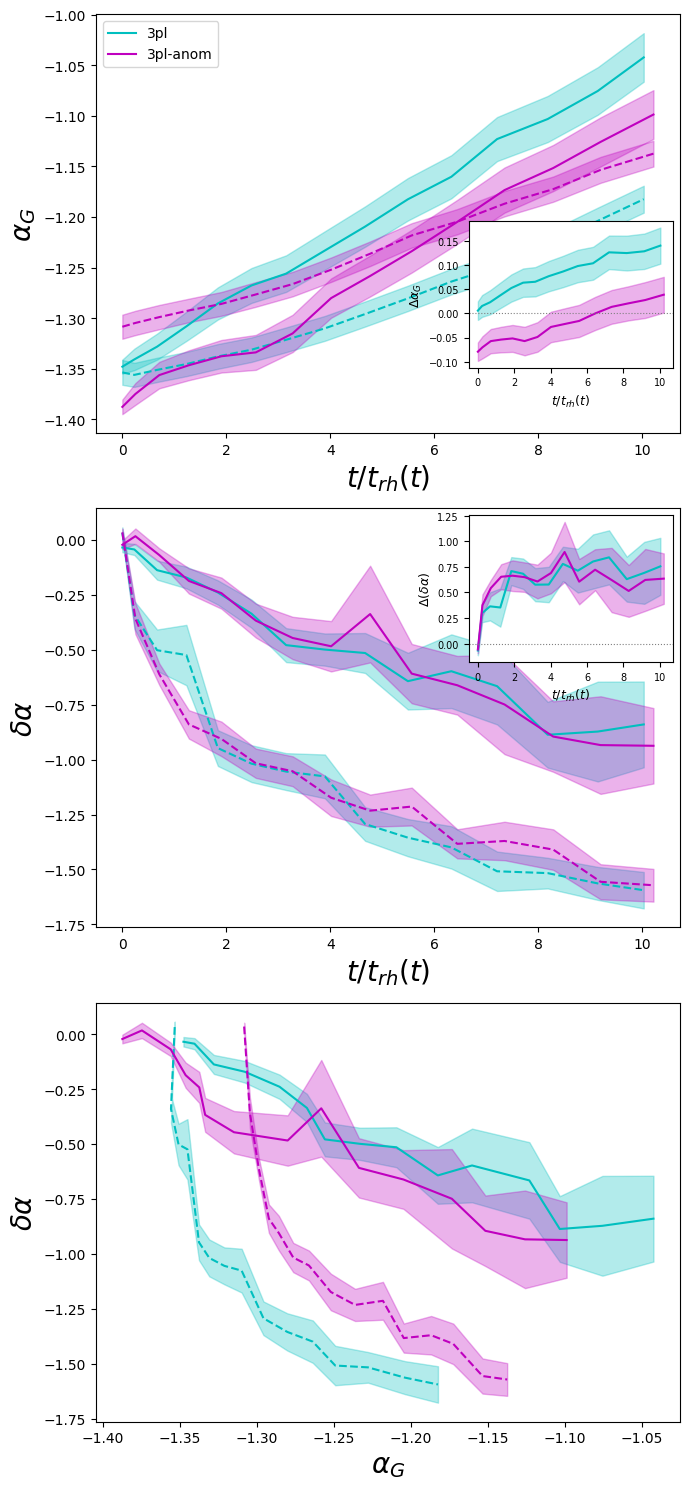}
    \caption{Same as Figure~\ref{fig:fig 18}, but within the mass range $0.3$--$0.8\,M_\odot$.}
    \label{fig:fig M10}
\end{figure}

\begin{figure}[h]
    \centering
    \includegraphics[width=0.95\linewidth]{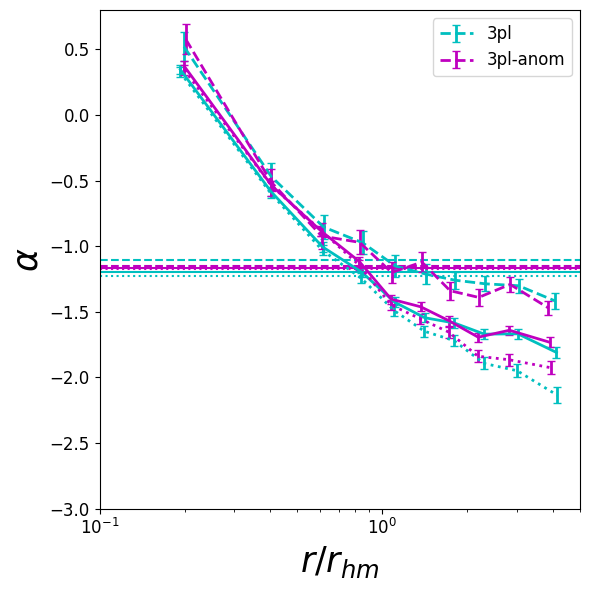}
    \caption{Same as Figure~\ref{fig:fig 16}, but within the mass range $0.3$--$0.8\,M_\odot$.}
    \label{fig:fig M11}
\end{figure}

\begin{figure}[h]
    \centering
    \includegraphics[width=0.95\linewidth]{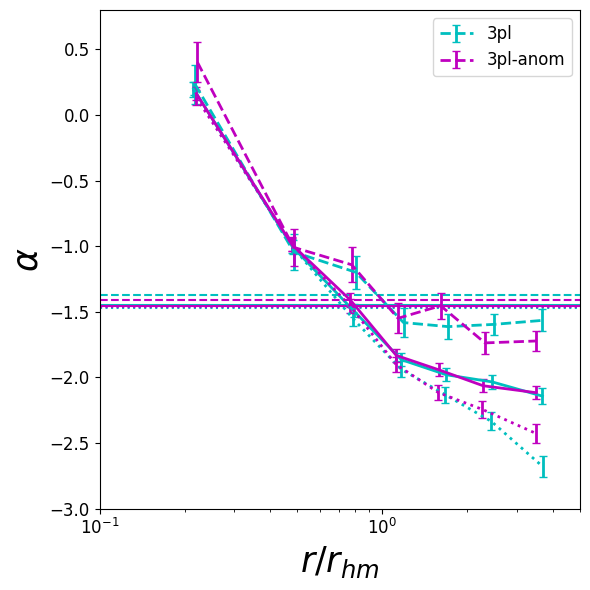}
    \caption{Same as Figure~\ref{fig:fig 16}, but within the mass range $0.4$--$0.8\,M_\odot$.}
    \label{fig:fig M12}
\end{figure}

\begin{figure}[h]
    \centering
    \includegraphics[width=0.95\linewidth]{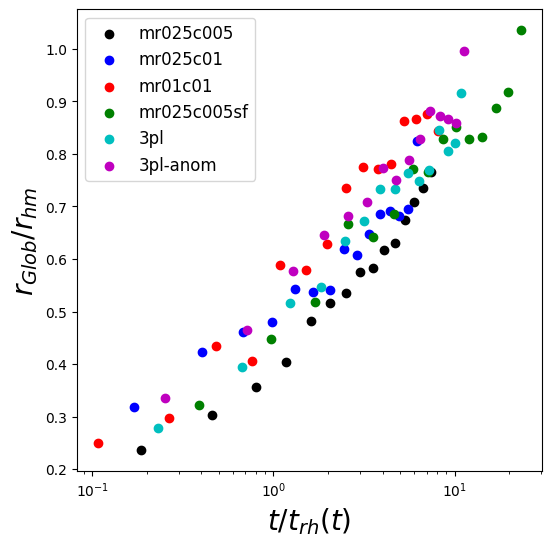}
    \caption{Same as Figure~\ref{fig:fig 11}, but within the mass range $0.3$--$0.8\,M_\odot$.}
    \label{fig:fig M13}
\end{figure}

\begin{figure}[h]
    \centering
    \includegraphics[width=0.95\linewidth]{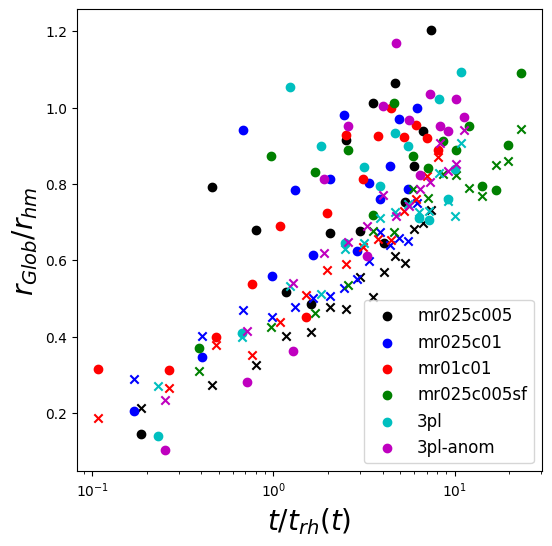}
    \caption{Same as Figure~\ref{fig:fig 12}, but within the mass range $0.3$--$0.8\,M_\odot$.}
    \label{fig:fig M14}
\end{figure}

\begin{figure}[h]
    \centering
    \includegraphics[width=0.95\linewidth]{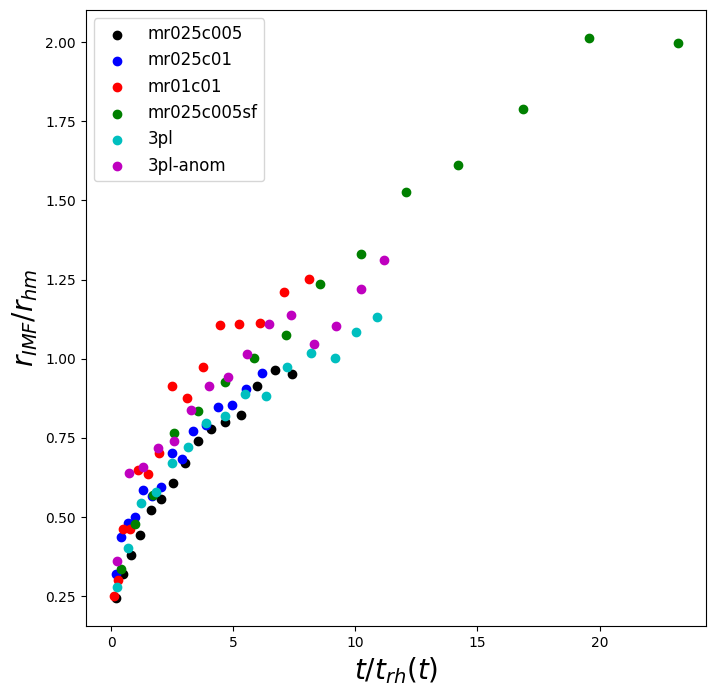}
    \caption{Same as Figure~\ref{fig:fig 13}, but within the mass range $0.3$--$0.8\,M_\odot$.}
    \label{fig:fig M15}
\end{figure}

\begin{figure}[h]
    \centering
    \includegraphics[width=0.95\linewidth]{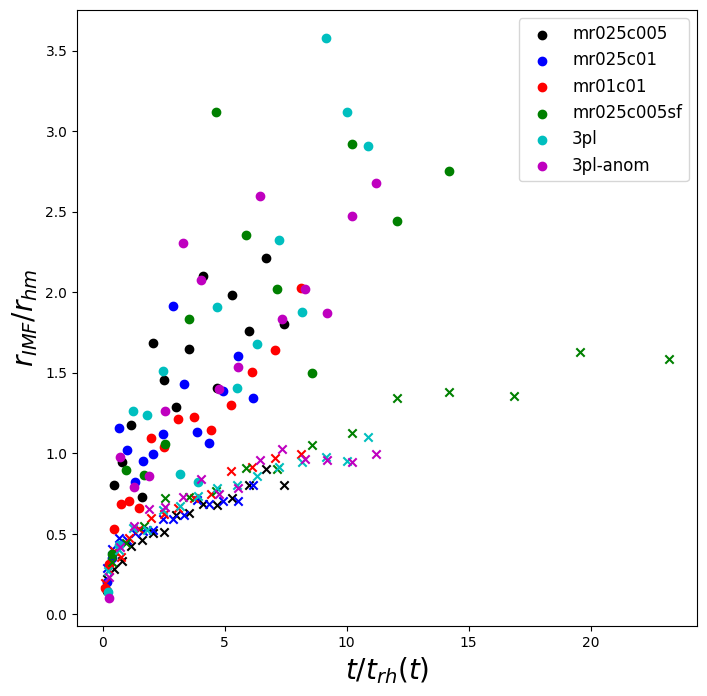}
    \caption{Same as Figure~\ref{fig:fig 14}, but within the mass range $0.3$--$0.8\,M_\odot$.}
    \label{fig:fig M16}
\end{figure}

\section{Observational Projections}
\label{sec:A2}

In this section we present results using projected 2D coordinates, showing how the intrinsic 3D radial distributions and characteristic locations described in the main text could be compared with observational results. For each snapshot, we generate a random 2D projection by sampling a random orientation for the line of sight and computing the projected distance of each star from the cluster center. Using multiple realizations per snapshot had a negligible impact on the resulting profiles (smaller than the uncertainties in $\alpha$), and we compute the profiles normalized to the projected half-light radius. Here we report the values within the mass range $0.1$--$0.5\,M_\odot$ (K01) or $0.1$--$0.4\,M_\odot$ (3pl).

Figure~\ref{fig:fig M18} and \ref{fig:fig M19} show the 2D projected radial variation in the MF slope at 12 Gyr for the total cluster and within each stellar population, respectively. Figures~\ref{fig:fig M20} and \ref{fig:fig M21} show the time evolution of $R_{Glob}$ for the total cluster and within the multiple stellar populations. Figures~\ref{fig:fig M22} and \ref{fig:fig M23} show the corresponding time evolution of $R_{IMF}$. Finally, Figure~\ref{fig:fig M24} shows the 2D projected radial profiles for the 3pl and 3pl-anom simulations split into their multiple populations.

\begin{figure}[h]
    \centering
    \includegraphics[width=0.95\linewidth]{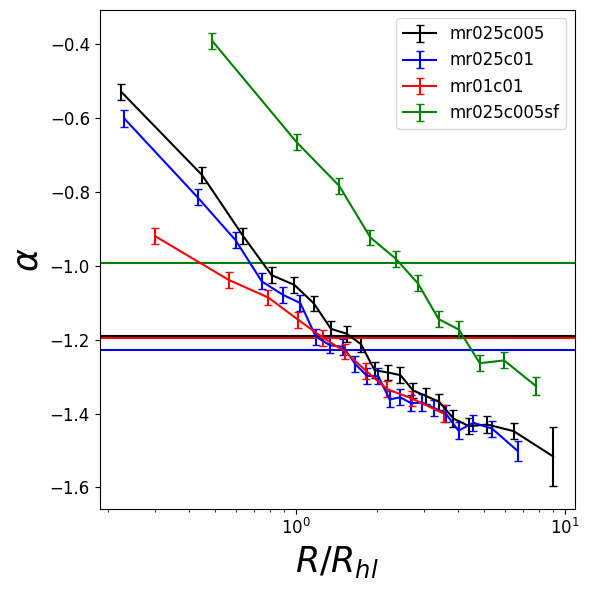}
    \caption{Same as the top panel of Figure~\ref{fig:fig 3}, but in 2D projected coordinates, normalized to the 2D projected half-light radius.}
    \label{fig:fig M18}
\end{figure}

\begin{figure}[h]
    \centering
    \includegraphics[width=0.95\linewidth]{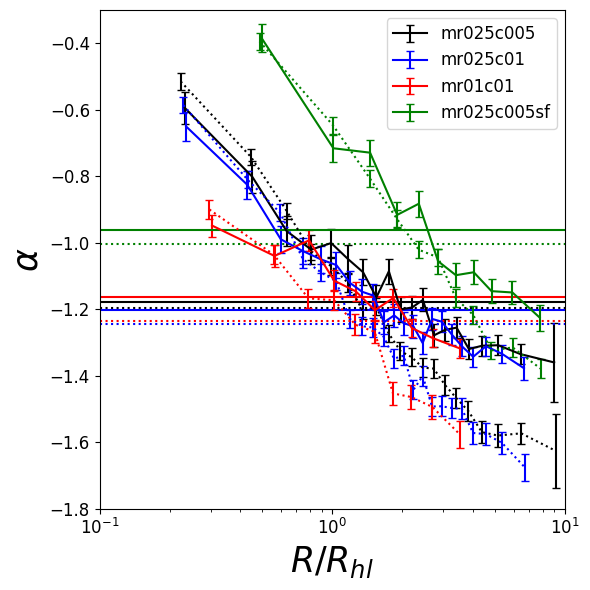}
    \caption{Same as the top panel of Figure~\ref{fig:fig 4}, but in 2D projected coordinates, normalized to the 2D projected half-light radius.}
    \label{fig:fig M19}
\end{figure}

\begin{figure}[h]
    \centering
    \includegraphics[width=0.95\linewidth]{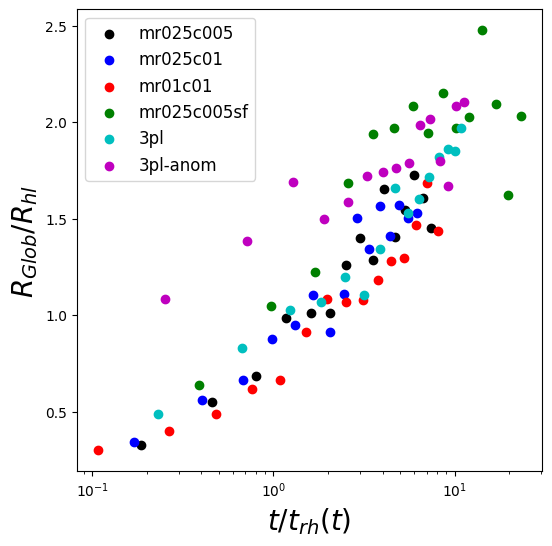}
    \caption{Same as Figure~\ref{fig:fig 11}, but in 2D projected coordinates, normalized to the 2D projected half-light radius. }
    \label{fig:fig M20}
\end{figure}

\begin{figure}[h]
    \centering
    \includegraphics[width=0.95\linewidth]{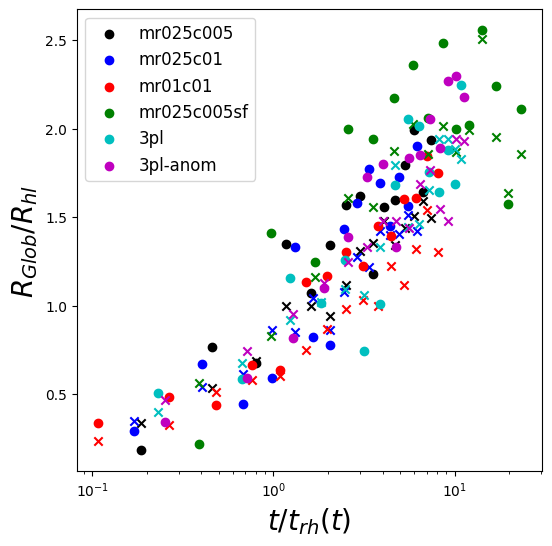}
    \caption{Same as Figure~\ref{fig:fig 12}, but in 2D projected coordinates, normalized to the 2D projected half-light radius. }
    \label{fig:fig M21}
\end{figure}

\begin{figure}[h]
    \centering
    \includegraphics[width=0.95\linewidth]{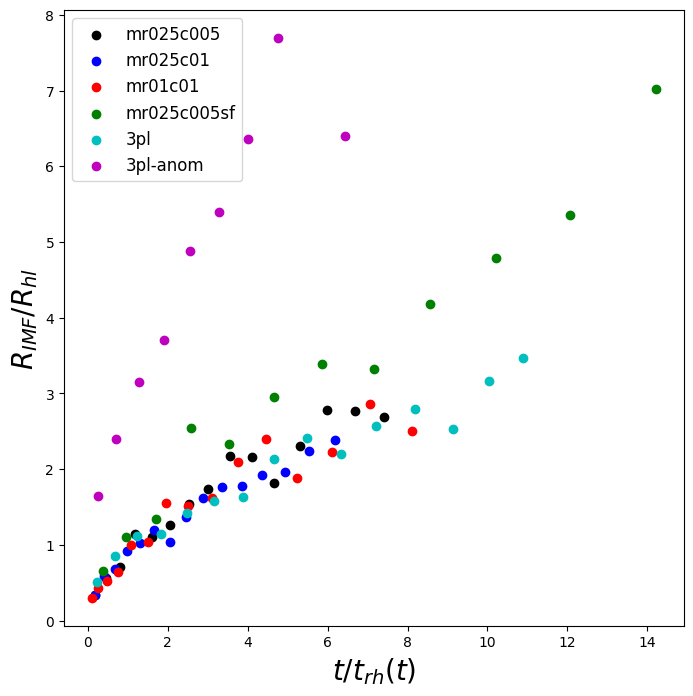}
    \caption{Same as Figure~\ref{fig:fig 13}, but in 2D projected coordinates, normalized to the 2D projected half-light radius. }
    \label{fig:fig M22}
\end{figure}

\begin{figure}[h]
    \centering
    \includegraphics[width=0.95\linewidth]{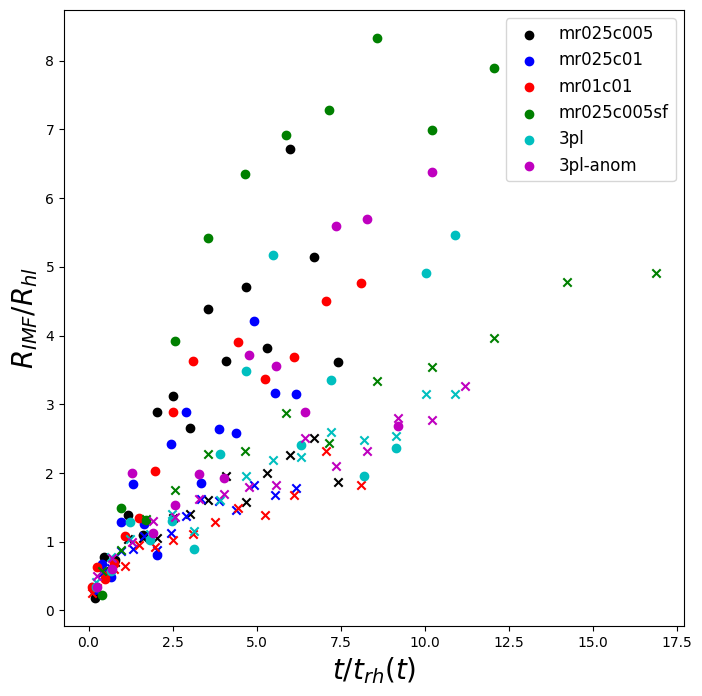}
    \caption{Same as Figure~\ref{fig:fig 14}, but in 2D projected coordinates, normalized to the 2D projected half-light radius. }
    \label{fig:fig M23}
\end{figure}

\begin{figure}[h]
    \centering
    \includegraphics[width=0.95\linewidth]{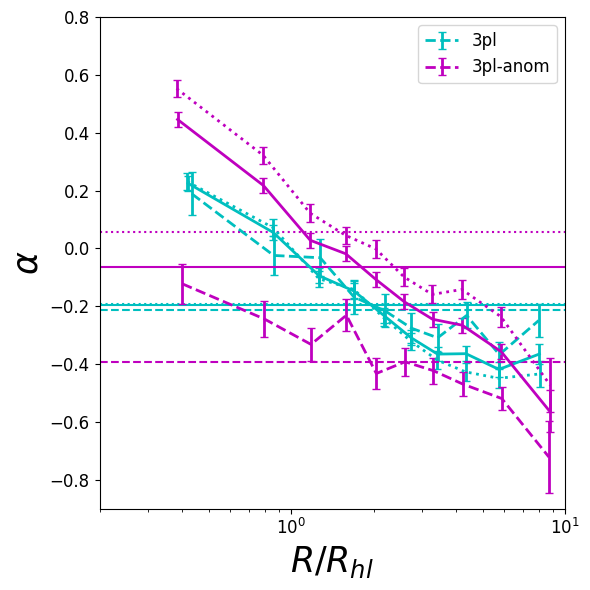}
    \caption{Same as Figure~\ref{fig:fig 16}, but in 2D projected coordinates, normalized to the 2D projected half-light radius. }
    \label{fig:fig M24}
\end{figure}

\end{appendix}

\end{document}